\documentclass[compsoc,conference,a4paper,10pt,times]{IEEEtran}
\IEEEoverridecommandlockouts
\usepackage{cite}
\usepackage{amsmath,amssymb,amsfonts}
\usepackage{graphicx}
\usepackage{textcomp}
\usepackage{xcolor}
\usepackage[colorlinks=true,urlcolor=black]{hyperref}

\usepackage[utf8]{inputenc}
\usepackage{epsfig}
\usepackage{verbatim}
\usepackage{subcaption}
\usepackage{times}
\usepackage{color, colortbl}
\usepackage{amsthm}
\usepackage{tikz}
\usepackage{url}
\usepackage{comment}
\usetikzlibrary{arrows.meta}

\def\BibTeX{{\rm B\kern-.05em{\sc i\kern-.025em b}\kern-.08em
    T\kern-.1667em\lower.7ex\hbox{E}\kern-.125emX}}

\newcommand{\claudia}[1]{\textcolor{violet}{[cd: #1]}}
\newcommand{\aggelos}[1]{\textcolor{blue}{[ak: #1]}}
\newcommand{\harry}[1]{\textcolor{teal}{[hh: #1]}}
\newcommand{\ignore}[1]{}

\newcommand{\vrfK}{\mathsf{K}} 
\newcommand{\vrfE}{\mathsf{E}} 
\newcommand{\vrfV}{\mathsf{V}} 

\newtheorem{theorem}{Theorem} 
\newtheorem{definition}{Definition}

\begin{document}

\title{Decentralized Reliability Estimation for Low-Latency Mixnets}

\title{Decentralized Reliability Estimation for Low-Latency Mixnets\\
\thanks{This research was supported in part by
CyberSecurity Research Flanders with reference number
VR20192203 and by Nym Technologies SA.}
}

%%%%%%%%%%%%%%%% Authors' Info %%%%%%%%%%%%%%%%%
%%

\author{\IEEEauthorblockN{ {
Claudia Diaz}}
\IEEEauthorblockA{\textit{Nym Technologies SA and KU Leuven} \\
Leuven, Belgium \\
claudia.diaz@esat.kuleuven.be}
\and
\IEEEauthorblockN{Harry Halpin}
\IEEEauthorblockA{\textit{Nym Technologies SA} \\
Neuch\^{a}tel, Switzerland \\
harry@nymtech.net}
\and
\IEEEauthorblockN{ {
Aggelos Kiayias}}
\IEEEauthorblockA{\textit{University of Edinburgh and IOG} \\
Edinburgh, UK \\
aggelos.kiayias@ed.ac.uk}
\IEEEpeerreviewmaketitle
}

\maketitle
%\color{red}
%\textbf{Preprint accepted to EuroS\&P 2026, to be replaced by the camera-ready version by end April 2026}
%\vspace{0.1cm}
%\\  

\color{black}
\begin{abstract} 

Mix networks (\textit{mixnets}) enable anonymous packet-based communication via multi-hop routing. Measuring the reliability of mix nodes serving as intermediate hops in a way that is decentralized, accurate, public, and compatible with low-latency packet routing, remains, however, a challenge: % -- despite being essential for maintaining a usable service. 
existing \textit{verifiable mixing} schemes introduce routing delays ranging from minutes to hours that grow with traffic volume, severely limiting their practicality for large-scale or latency-sensitive applications. 
We propose a decentralized scheme that provides public reliability estimates for a mixnet's links and nodes without increasing latency for client traffic. The scheme achieves optimal complexity, independent of total network traffic, and enables accurate, low-overhead estimation even at scale.
Our approach relies on verifiable measurement packets generated via a novel \textit{VRF-based routing} primitive. This mechanism produces unforgeable measurement packets indistinguishable from normal traffic, while ensuring that all packet routing choices are consistent with the mixnet’s routing policy. Reliability scores are derived from revealed measurement packets.
The scheme remains robust under both unreliable and adversarial conditions, producing accurate reliability estimates as long as every honest node has a majority of honest predecessors and successors in the routing graph. 
We validate our design experimentally, demonstrating its practicality and effectiveness.

\end{abstract}


  %% old content saved in sections/z-OLD-s1-intro.tex and in old_main-2023-01-17.tex

\section{Introduction}

A practical anonymous communication infrastructure must not only protect users’ privacy but also provide a dependable service. Anonymity networks such as Tor\footnote{\url{https://www.torproject.org}} and Nym\footnote{\url{https://nym.com/mixnet}} route user traffic through multiple intermediary nodes to protect privacy in the face of network observers and malicious infrastructure~\cite{dingledine2004tor, diaz2021nym}. However, real-world deployments inevitably include unreliable nodes—due to misconfiguration, poor connectivity, transient failures, or adversarial behaviour. Node reliability directly affects usability: if too many nodes fail too often, the network degrades and may become unusable.

This is particularly important in mixnets such as Nym,
%which take a {\em privacy for clients} but {\em transparency and accountability for the infrastructure} approach:  \aggelos{Claudia: check}
whose node selection and incentive mechanisms require accurate reliability estimates~\cite{Diaz2022Reward-mixnet}. A node’s reliability score determines both its likelihood of being selected to route user traffic in the next \textit{epoch} and the rewards it receives for its service. Nym nodes are thus incentivized to maximize reliability to compete for participation and financial rewards.
Unlike Tor, where circuit-based connections allow clients to directly observe relay reliability~\cite{greubel2020, zhang2021tnras}, mixnets use stateless, per-packet routing. Each packet is routed independently, which complicates the task of determining where losses occur along a multi-hop path.

Prior work on mixnet reliability typically takes an ``all-or-nothing'' approach 
that equates packet-loss with malicious behaviour: the objective is not, as in our case, reliability estimation scoring that estimates the level of packet loss at each component, but rather, to identify the ``bad nodes'' that dropped one or more packets, so they can be excluded.   
One dominant approach to ensure correctness assumes public broadcast of all packets,
as well as verifiable shuffles, at each hop in the communication path. 
These systems offer strong integrity and verifiability but they incur significant latency for all traffic, making them unsuitable for low-latency applications. Voting systems are a common use case~\cite{DBLP:conf/fc/FurukawaS06,DBLP:conf/ccs/Neff01}, and more efficient designs~\cite{kwon2017atom, tyagi2017stadium, XRD} have extended this model to messaging. Still, even these faster systems impose routing delays of several minutes, which scale with traffic volume, rendering them ill-suited for general-purpose internet access.
On the more efficient side, tripwires \cite{DBLP:conf/asiacrypt/KhazaeiMW12,DBLP:conf/esorics/BoyenHM20} also enable identifying nodes that drop packets, but again are not designed for producing node reliability scores. 

Supporting reliability in a low-latency mixnet thus requires new techniques that preserve privacy and performance at scale while also offering an accurate \textit{numerical} estimation of the reliability of participants. %These must function in a continuous-time, packet-based setting, without relying on centralized components.
A low-latency mixnet like Nym, which routes packets with sub-second latency,\footnote{Nym (\url{https://nym.com/explorer}) currently routes packets with $500$ms to $800$ms end-to-end latency, including mixing and propagation. Packet processing itself takes under $1$ms and is negligible by comparison.} is incompatible with existing verifiable routing mechanisms. While end-to-end acknowledgments and retransmissions can improve reliability~\cite{piotrowska2017loopix}, they do not identify which mix node caused a failure, nor do they provide public, network-wide reliability estimates—data that is essential for maintaining quality of service, managing the network, and incentivizing node behaviour~\cite{Diaz2022Reward-mixnet}.
This gap points to a key challenge: to make mixnets viable for a broad range of applications, they must combine reliability estimation with low-latency operation--—a combination that has remained elusive. Moreover, packet delivery latency should remain independent of the total traffic volume to preserve performance as the network scales. Finally, reliability estimation must be decentralized and publicly verifiable to avoid trusted components becoming single points of failure.\footnote{Due to the lack of low-latency decentralized solutions, the Nym network currently relies on centralized \textit{Network Monitors} that send loops of traffic through the mixnet to derive node reliability scores.}

%The second approach is \emph{continuous time mixnets} that offer the possibility of bounding routing latency by having mix nodes apply random delays to packets, such that the latency of routing packets is independent of the number of packets~\cite{kesdogan1998stop, piotrowska2017loopix}. 

%Mix networks or \textit{mixnets}, first introduced by Chaum~\cite{chaum1981untraceable}, are a technique for anonymously routing packets through multiple intermediaries called \emph{mix nodes}, such that the input and output packets of nodes are unlinkable due to cryptographic transformations and packet reordering (i.e. mixing). Mixnets route packets independently,
%each one paving its own path through the mixnet.  This is in sharp contrast to connection-based approaches like onion routing~\cite{onion-routing-1996} as used in Tor~\cite{dingledine2004tor}, which establish end-to-end bidirectional circuits carrying all the packets of a communication flow --- in this way providing the initiator with a live connection to each routing intermediary that can estimate relays' reliability directly. 

%Mixnet design follows two major approaches. The first is \textit{batch-based}~\cite{chaum1981untraceable, chaum2017cmix, kwon2017atom} mixnets that collect and then shuffle batches of $N$ packets. Batch-based 

\noindent 
{\bf Our contributions.} 
We propose a decentralized reliability estimation protocol that is compatible with practical low-latency mixnets, as it preserves sub-second
packet latency and has an overhead that remains constant independently of packet payload size and overall volume of client traffic being routed. 
Our scheme achieves this at a fraction of the overhead of verifiable shuffling techniques, while still relying on broadcast to achieve public verifiability.
It is designed to integrate with Nym’s credential and incentive systems~\cite{DBLP:journals/popets/RialP23, Diaz2022Reward-mixnet, diaz2021nym}, but remains generic and analyzable as a stand-alone protocol.

The intuition behind our solution is that of an army of undercover ``secret shoppers'' who visit a store in a manner representative to the overall shopping patterns and aim to detect  faulty behavior in employees. This idea expands the concept of tripwires and can be realized through a novel primitive that we call \textit{VRF-based routing}. 
A VRF (Verifiable Random Function)~\cite{DBLP:conf/focs/MicaliRV99} %is a function that  
maps any input to a pseudorandom value in a way that is publicly verifiable.
In our setting, VRFs ensure that packet routes are randomly sampled from the mixnet’s publicly known \textit{routing policy}. 
%\claudia{is the following sentence still true?? verify} 
%\aggelos{no -  commenting it out } 
%The primitive additionally includes random packet legitimacy challenges that ensure that malicious entry points who allow non-legitimate packets into the network are publicly exposed, while maintaining packet information private. 
VRF-based routing allows turning a mixnet's packet encoding scheme (such as the widely deployed Sphinx packet format~\cite{sphinx-2009}) into a verifiably unbiased lottery designating a subset of packets as \textit{measurement packets}.
These measurement packets are used for network monitoring rather than payload delivery, yet remain indistinguishable from normal traffic while in transit. At a later stage, they can be selectively revealed and verified to support reliability estimation.
Building on this mechanism, we design a protocol that periodically produces public estimates of the fraction of traffic transmitted or dropped on each mixnet link during a time epoch. From these link-level statistics, we further derive \textit{node} reliability numerical 
scores based on the pattern of dropped packets on their incoming and outgoing links.

To understand the power of the approach, consider a communication link involved in $n$ packet transmissions, exhibiting some arbitrary reliability pattern, correctly routing packets with unknown probability $\rho$ and dropping them with probability $1-\rho$.  Using VRF-based routing we can seamlessly embed a series of indistinguishable measurement packets within the traffic. 
Because these packets are indistinguishable from normal traffic, the link can be viewed as a lossy channel that behaves like a biased coin with unknown bias $\rho$. The opening of the commitments to measurement packets yields a publicly observable sequence of coin flips from this $\rho$-biased coin.
From this sequence, the conditional distribution of $\rho$ follows a Beta distribution with variance $O(1/t)$, where $t$ is the number of measurement packets. It follows that we can estimate $\rho$ up to error $\sim 1/\sqrt{t}$ with broadcast communication complexity equal to $O(t)$. 
Note that this overhead is independent of the total amount $n$ of client packets sent in the link, and depends only on the amount $t$ of measurements per link. The sample complexity is thus essentially optimal, as this number of coin flips is required to estimate a bias with this level of accuracy.

\ignore{%%%  As above but via Hoeffding bounds 
To understand the power of the approach consider a communication link involved in $n$ packet transmissions, exhibiting some arbitrary reliability pattern denoted by $b_i, i=1,\ldots, n$, with $b_i = 1$ iff the $i$-th packet was successfully sent via the link. VRF-based routing enables to transform the $i$-th packet into an indistinguishable measurement packet with probability $p_\mathsf{lot}$, say $X_i=1$ iff $i$-th packet is a measurement. By revealing the measurements  
one can estimate reliability by $T = \sum^n_{i=1} b_i X_i$. In more detail, 
if $\mu$ is the mean of $T$, observe that $\mu = pt$, where $p$ is the measurement packet lottery probability
and $t = \sum^n_{i=1} b_i$.
Applying Hoeffding's inequality we have $\mathbf{Pr}[|T - \mu| \geq \epsilon \mu ] \leq 2\exp(-(\epsilon\mu)^2 )\leq \delta$. %, from which we conclude that $T$ estimates $pt$  within error $\epsilon$ with confidence $(1-\delta)$  as  long as $ pt \geq \epsilon^{-1}\ln^{1/2}(\delta^{-1}/2)$. 
Now if
$\chi = t/n$ is the true average reliability of the relay,  
by setting $p = k/n$ where $k$ is a security parameter, we obtain that $T/k$ estimates $\chi$ with error $\epsilon$ and with confidence $(1-\delta)$ as long as $k$ is large enough, specifically $k\geq  (\chi\epsilon)^{-1}\ln^{1/2}(\delta^{-1}/2)$. The critical benefit here is that we obtain a good estimation of true reliability with only $pn = k$ measurements where $k$ is a security parameter that is {\em independent} of the number of total packets $n$, which may be very large. 
}%%%%%

We empirically evaluate the proposed protocols through simulations. Simulation offers the advantage of knowing the ground truth --- that is, which nodes are actually unreliable --- allowing us to directly assess how accurately our protocols approximate this ground truth. This level of validation is not possible in real-world deployments, where ground truth is inherently unavailable.
Our results show that, given sufficient measurement samples, the node reliability estimation protocol produces accurate estimates of failure rates in scenarios where nodes are honest but unreliable. Furthermore, we demonstrate that malicious nodes attempting to reduce their neighbours’ reliability scores by selectively dropping packets incur a reliability penalty equivalent to the harm they inflict. This protects against “creeping death” attacks~\cite{dingledine2003reliable}, where adversaries selectively drop packets to gradually gain reputational advantage over honest nodes.
Our protocols are highly scalable: the number of required measurements and the associated overheads remain \textit{constant} for a given estimation accuracy and network topology (specifically, number of links in the network), rather than growing with the overall volume of client traffic or with the packet size, as is the case with existing solutions.

\noindent
{\bf Paper organization.}
Section~\ref{sec:problem} introduces the problem statement, including the system model, threat model, and desired properties. Section~\ref{sec:sampling-protocol} presents our proposed protocol: we first describe the estimation procedure for mixnet link reliability, then introduce the VRF-based routing primitive that underpins our approach, and finally explain how to compute reliability scores for mix nodes. Section~\ref{sec:empirical-evaluation} evaluates the accuracy of the proposed methods through simulations. Section~\ref{sec:related} discusses related work, and Section~\ref{sec:conclusion} concludes.

\section{Model and Problem Statement}

%In this section, we define the elements of the mixnet system used, the threat model, and then formally define mixnets with reliability estimation.
\label{sec:problem}
\color{black}

%\claudia{@aggelos I would update this section to contain: system model (including assumed building blocks such as broadcast channel, incentive system, credential system, etc.), threat model, and desired properties (eg: measurements cannot be fabricated, biased, or distinguished; everyone (decentralized) can estimate drop rate per link and reliability score per node; adaptively dropping packets causes symmetric cost to victim and attacker); ie, move here parts of what is now Sect. 4. Then Introduce VRF routing in the next Section 3, and finally merge the remains of section 4 with 5 for the reliability estimation protocol.}

\subsection{System Model}

\noindent 
{\bf Decryption mixnet model.}
We consider a %continuous-time \textit{decryption mixnet} 
mixnet where senders prepare packets by encrypting them with the keys of selected intermediaries. Each intermediary removes one layer of encryption when routing the packet. All packets are assumed to have the same length. We model the mixnet as a labeled directed graph $G$, whose vertices represent \textit{mix nodes}, \textit{gateways}, and \textit{clients}, and edges represent communication links. We denote the edge from vertex $i$ to $j$ as $e= (i,j)$. 
Mix nodes have both incoming and outgoing edges to other mix nodes and gateways. 
Gateways serve as both entry and exit points for the mixnet, with links to both clients and mix nodes. While not part of some mixnet designs, gateways are standard for Loopix-based designs, where they are called "providers", to support reliability and be able to receive replies~\cite{piotrowska2017loopix}.
%\claudia{perhaps here introduce the notion of packet "route" as a path of edges in $G$ from client sender/source to client receiver/destination, chosen by the sender according to the following?} 
For each entity $j$, we denote as $P(j)$ and $S(j)$, respectively, the sets of its immediate predecessors and
successors in $G$. 
%The set of \emph{forward direction predecessors} denoted by $P^*(j)$ is defined to be $\emptyset$ for a client $j$, a set of clients in case $j$ is a gateway, and  
%P^*(j) = P(j) \cup_{i \in P^*(j)} P^*(i)$ for a mix node $j$. 
%As illustrated in Fig.~\ref{fig:overall-system} (steps A.2 to A.6), 
A packet route is initiated by a client  affiliated with a gateway $g_0$ and follows a path in $G$ picking outgoing edges and traversing a sequence of mix nodes. The last mix node is followed by the recipient's gateway $g_1$, which delivers the packet to its intended destination (e.g., a client, a service, or a proxy to the open internet). 
%This is illustrated in Fig.~\ref{fig:overall-system} (steps A.2-A.6). 
%\claudia{@aggelos: should we say anything here about "source routing" and each intermediary "decrypting" a layer of the packets? the headline says 'decryption mixnet model' but we only talk about the graph, nothing about "decryption". Added some lines at the start to address this.}
%Once processed at $n_\mathsf{exit}$ the packet will be sent to its intended destination, which could be another client. 
%\claudia{...or? do we want to mention alternatives?}\aggelos{let's decide later based on space limitations}.  

%\claudia{should we have a small figure to illustrate the graph G?} \aggelos{yes, provided space permits..}

%\harry{Users buy a {\em subscription} by performing a payment on a smart contract on the blockchain. This subscription has the form of a compact e-cash wallet~\cite{DBLP:journals/popets/RialP23} (with keys $vk_\mathsf{long},  sk_\mathsf{long}$) that contains a bundle of $N$ unlinkable \textit{credentials}. 
%\claudia{and is registered on chain?? @@aggelos --- we may want to provide some more info here + ref to key notations in section 3}. 
%Each credential $c$ entitles the client to route up to $S_c$ packets through the mixnet. }

%The credential $c$ is also registered on chain. 

\noindent
{\bf Client credentials.}
Users obtain a bundle of unlinkable \textit{credentials}, e.g., in the form of a compact e-cash wallet~\cite{DBLP:journals/popets/RialP23}, 
with verification and spending keys $vk_\mathsf{long}$ and $sk_\mathsf{long}$, respectively. 
Each credential $c$ authorizes a client to send up to $S_c$ packets through the mixnet via a gateway of their choice, where $S_c$ is a predetermined value. This is the same design Nym uses to authenticate users, and similar schemes are used for example by PrivacyPass in Tor~\cite{davidson2018privacy}. 

\noindent 
{\bf Network topology and routing policy.}
We assume for convenience that the mixnet graph $G$ has a layered topology, i.e., $N=LW$ mix nodes are arranged in $L$ layers, each containing $W$ nodes.
%(we assume each layer is the same size in our analysis for sake of simplicity). 
Mix nodes are drawn from a pool of at least $N$ eligible nodes and are randomly assigned to layers at the start of each \textit{epoch}. Assignments are refreshed every epoch to rotate node positions and allow replacement of underperforming nodes. A set of $W_G$ gateways is assumed to remain stable across epochs to minimize the need for clients to switch gateways with the change of epoch. 
Clients spend credentials with eligible gateways to send traffic to the mixnet, as first-layer nodes only accept packets from gateways. Valid packet routes traverse one node per layer, with each hop chosen independently and uniformly at random among the $W$ nodes in the layer. Mix nodes in the last mixnet layer forward packets to their destination gateways.

\noindent 
{\bf Blockchain and broadcast channel.} 
We assume the presence of a blockchain that stores transactions persistently and can be used to post public commitments. Additionally, a ``layer 2'' ephemeral broadcast channel enables one-to-many dissemination of information that need only be available temporarily (e.g., for several hours). 
We refer to publishing on this ephemeral layer as ``broadcasting.'' 
Finally, the blockchain can be used to draw publicly verifiable randomness that all parties can receive, while being unknown to the adversary in advance.% i.e., it offers random \textit{beacon} values that are both unpredictable and publicly verifiable.

%%%%%%%%%%%%%

\subsection{Threat Model}
\label{sec:threat}

The proposed reliability estimation mechanisms produce outputs that support both network management and incentive allocation, such as determining node selection for mixnet service provisioning and assigning financial rewards. The primary goals are to ensure that the resulting measurements are \textit{accurate}, i.e., they approximate ground truth closely, and can support the incentive mechanism, meaning that nodes are motivated to maximize their reliability in order to increase their \textit{utility}. In the Nym system, utility corresponds to financial rewards and participation opportunities, but it can also take the form of \textit{reputation} in other networks~\cite{dingledine2001reputation, dingledine2003reliable}.

We assume an adversary that controls a subset of mix nodes and attempts to bias the reliability estimates, either to improve its own scores or to degrade those of others, without incurring a utility loss. In particular, the adversary may drop packets or otherwise degrade service in ways intended to shift performance metrics. In an ideal outcome, reliability estimates are robust against such manipulation, and adversarial nodes are best off maximizing their reliability, as any deviation leads to lower utility.

Because our central objective is to produce \textit{accurate} and \textit{unbiased} reliability estimates, we focus on adversarial strategies that distort measurement outcomes beyond the expected error margins of the system.  
We note that in realistic deployments, even high-quality, well-behaved nodes may experience occasional failures. As such, it is impractical to treat every packet loss as evidence of malicious behaviour~\cite{miranda}. Instead, we assign continuous reliability scores in the range [0,1], allowing the system (i) to preferentially select high-performing nodes at each epoch, and (ii) to incentivize all participants to maintain the highest possible reliability over time.

From a privacy perspective, it is essential that the proposed protocols do not reveal any \textit{new information} that could be exploited for privacy attacks. We acknowledge that reliability degradation may, in some scenarios, serve as a stepping stone for broader privacy compromises~\cite{borisov2007denial}. Nonetheless, adversaries whose primary goal is to compromise privacy are considered out of scope for this work, which focuses on the integrity and robustness of reliability estimation.

The threat model considered here relies on the following assumptions:

%\claudia{@aggelos: should we explicitly say that we consider an "economically rational adversary"?? our schemes and analysis make more sense if the adversary's goal is to get some sort of economic advantage rather than to just vandalize a system for no gain. }

\noindent 
%$\bullet$ 
\textbf{Adversarially controlled entities.} 
The adversary may control any number of \textit{clients} and a limited number of nodes (\textit{mix nodes} and \textit{gateways}), under the conditions that every layer of the mixnet has at least one honest node and that at least half of the predecessors and half of the successors of every honest node in $G$ are also honest. 
Malicious nodes may adaptively drop or substitute packets and may broadcast incorrect output values. Clients under adversarial control may behave arbitrarily. Adversarial gateways are assumed to be utility-maximizing: they may follow any strategy that does not decrease their utility. The utility function is arbitrary but must be monotonically increasing in a set of indicators defined in Section~\ref{sec:vrf-based-routing}.

\noindent 
%$\bullet$ 
\textbf{Honest nodes.}
Honest nodes are divided into (i) \textit{reliable} nodes, which are consistently online throughout an epoch and forward all received packets, and (ii) \textit{unreliable} nodes that may experience downtime, throughput limitations, or other transient faults that result in dropped packets during the epoch. All honest nodes report correct output values in every epoch.

% (unless they are down). 
%\aggelos{Byzantine vs. faulty vs. unreliable vs. online  - what is the best terminology ? }  \claudia{I tend to use `malicious', `adversarial' or `actively adversarial' for those that make strategic choices to bias things in a certain way. For the others, i'm good with terms like `unreliable', `faulty' / `online', `reliable'. I will provide concrete info on types/variables to model "unreliable / faulty" and "malicious" in sect 4 (or appendix if no space)}
%\aggelos{ok let's settle for ``malicious, unreliable and reliable'' for respectively referring to adversarial nodes, honest but occasionally experiencing downtime, and honest and consistenly online respectively}. 

%\item A demand distribution\claudia{not sure we need a `distribution', or simply a \textit{counter} with the issued credential remaining credits, considering how much was spent in epochs since issuance until last epoch -- also are we using this anywhere? if not, not sure we need to include it here} that determines how many clients wish to use the mixnet and by how much in number of packets. Clients acquire credentials as needed.

\noindent 
%$\bullet$ 
\textbf{Credential issuer.} 
The issuer (or set of issuers) is assumed to issue credentials correctly but is not trusted with respect to client privacy or involved in measurement accuracy. 
% that issues credentials as requested but is not trusted with respect to the privacy of clients. %\claudia{we may need to explain a bit more what it's meant with the `privacy of the clients'} \aggelos{see definition above} 

%\claudia{@aggelos: should we explicitly say that we assume the blockchain and broadcast channels are not compromised and function as expected (broadcast information is available to all participants)?} \aggelos{agreed}

\noindent 
%$\bullet$ 
\textbf{Blockchain and broadcast channel.} 
We assume a blockchain and ephemeral broadcast channel that cannot be subverted by the adversary. Both satisfy consistency (honest parties eventually agree on their contents, possibly with delays) and liveness (messages posted eventually become available to all honest parties). 
%In the case of the ephemeral broadcast channel, unreliable honest parties may miss transmitted messages, while in the case of blockchain all honest parties eventually catch up on all posted transactions.
%Additionally, the blockchain provides regular, random \textit{beacon} values that are publicly verifiable and cannot be predicted in advance by the adversary. (repeated)

%\item The network links between mixnet nodes and gateways may be unreliable and can drop packets with some probability. ---> we explicitly assume this is not the case in most of the model, let's not include this here
%\claudia{I would include in the threat model: description of free riding behaviour, and description of malicious selective packet dropping to degrade honest nodes' reliability measurements}\aggelos{i think this would be redundant  - we have stated our properties above with some effort - so it should be clear now, that we are interested in preserving these properties given the presence of malicious clients}

\subsection{Defining Reliability Estimation}

\begin{figure}[ht]
    \centering
    \includegraphics[width=0.80\columnwidth]{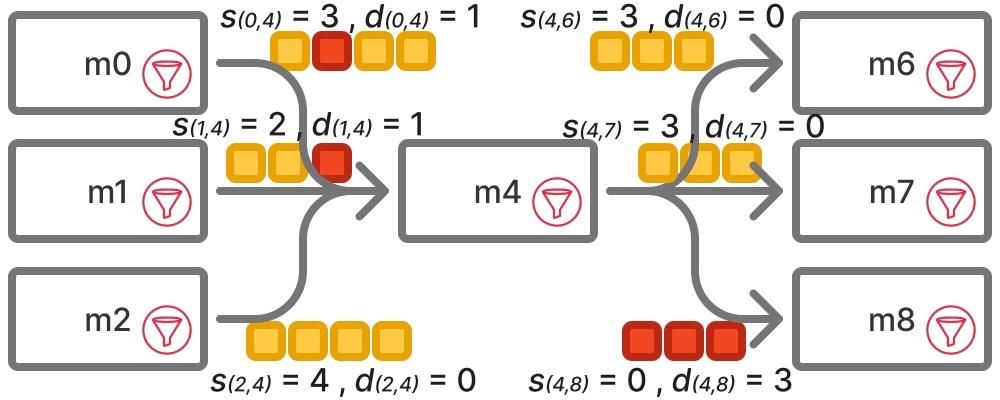}
    \caption{Example \textbf{${s}_e$} and \textbf{${d}_e$}  for the incoming and outgoing edges of node $m4$. Yellow packets are successfully transmitted, %on link $e$ %  $e = (i,j)$, %i.e., appearing in the tag-commitments of both $i$ and $j$; 
    while red packets are dropped. %, i.e., appearing in the tag-commitment of $i$ but not that of $j$.
    }
    \label{fig:link-traffic}
\end{figure}

The feature of \textit{reliability estimation} is introduced in addition to the usual mixnet operations for anonymously routing packets. 
A standard decryption mixnet includes two cryptographic algorithms: a \textit{packet encoding} algorithm that prepares content to be routed via the mixnet and a \textit{packet processing} algorithm that is executed at each hop in the packet route.  
%We illustrate in Fig.~\ref{fig:overall-system} the steps needed to anonymously deliver a packet in our model. 
Prior to running the packet encoding algorithm, we assume that clients have obtained %their e-cash wallet
%clients engage with an {\em issuer} %\harry{would prefer we normalize `voucher' to `anonymous credential'} 
%which provides them with an (anonymous divisible) 
credentials for using the system.
%\claudia{insert citation for Coconut?} \aggelos{no because this is not about coconut  - it is the voucher i am referring to}
They then approach a gateway of their choice and spend one of their credentials $c$, which entitles them to send $S_c$ packets through the mixnet.
Packets that originate from a given credential and gateway, and are correctly formed under the encoding algorithm, are called \textit{legitimate packets}. Note that a legitimate packet is well-defined given the gateway identity, the credential keys, and the random coins used by both client and gateway, assuming they follow the mixnet encoding algorithm.
%The client sends the packets to the gateway (step A.3), who, after checking for correctness, forwards them to the corresponding mix nodes (step A.4). These are further processed and forwarded by mix nodes (step A.5) and eventually delivered to the recipient (step A.6). 
%Given this feature, 
For each epoch, we associate the following two quantities with a link $e$ in the network graph $G$: 
\begin{itemize}
    \item \textbf{${s}_e$} is the number of legitimate packets that were successfully transmitted, i.e., sent and received across the link $e$.
    \item \textbf{${d}_e$} is the number of legitimate packets dropped in the link, i.e., reported as sent but never reported as received. 
\end{itemize}
Figure~\ref{fig:link-traffic} shows an example of \textbf{${s}_e$} and \textbf{${d}_e$} for the incoming and outgoing edges of a mix node $m4$. 
For client-gateway links we assume ${d}_{(c,g)}=0$, since these are live bidirectional links where clients may retransmit packets until the gateway acknowledges reception. Thus, ${s}_{(c,g)}$ includes all the packets submitted under credential $c$.\footnote{Note that each new credential $c$ allows the client to present a new identity towards gateways, thanks to the credentials' unlinkability properties.} 
At every subsequent hop, the quantities ${s}_e$ and ${d}_e$ of incoming and outgoing links of a gateway or node $j$ satisfy the following relation: 
%for any link $e$ that has $j$ as its endpoint. 
%(II) if  
 %$\chi_{e}$ denotes the fraction of time, when the  link $e$ drops a %packet either due to unreliable communication or due to the transmitting %counterparty being offline then we have:
%
%
\begin{equation} 
\sum_{i\in P(j)} {s}_{(i,j)} = \sum_{k\in S(j)}  
({s}_{(j,k)} +  {d}_{(j,k)})
\label{eq:sumpackets} 
\end{equation} 
where $\sum_{i} {s}_{(i,j)}$ is the number of packets received by $j$ from all its predecessors, $\sum_{k} {s}_{(j,k)}$ is the number of packets successfully transmitted to successors, and $\sum_{k} {d}_{(j,k)}$ is the number of packets received by $j$ but dropped before reaching their next destination. 
In the example of Fig.~\ref{fig:link-traffic}: $\sum_{i} {s}_{(i,4)}=9$, $\sum_{k} {s}_{(4,k)}=6$, $\sum_{k} {d}_{(4,k)}=3$. 
%This notion generalizes to longer paths in the graph, e.g., $ s_{(i,j,k)}$, denotes the number of packets sent from $i$ to $k$ routed via $j$. \claudia{<-- we don't need this anymore further in the paper (it was for free riding)} 

\begin{figure*}
\centering
    \includegraphics[width=0.90\textwidth]{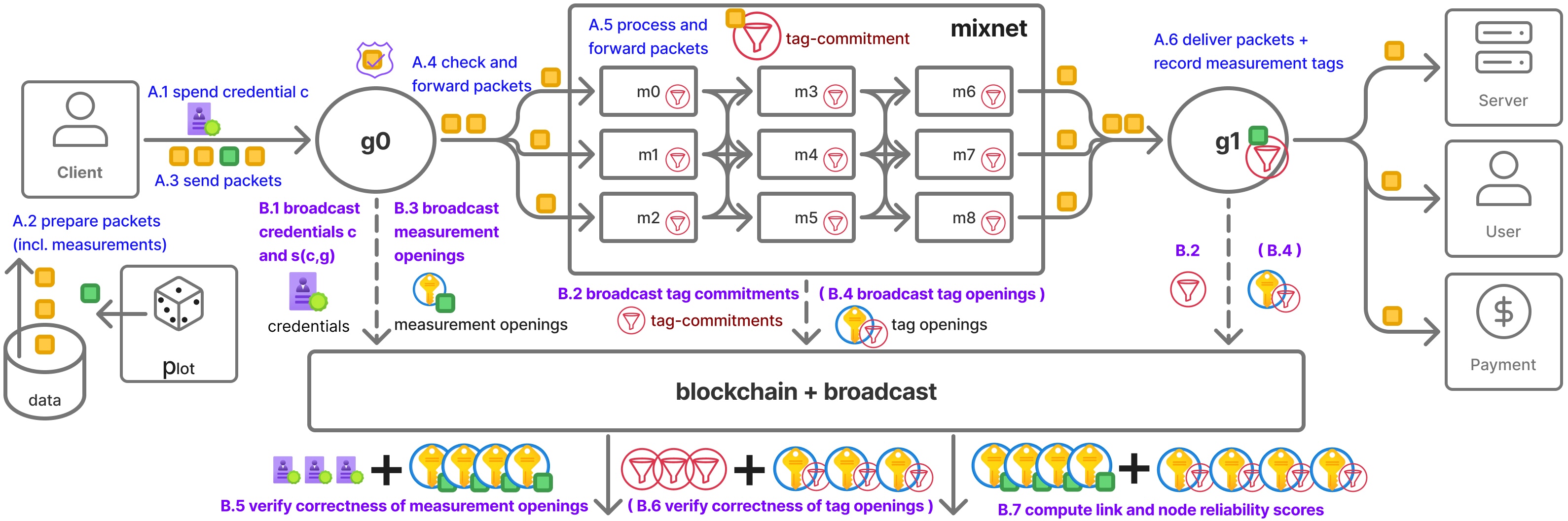}
\caption{\label{fig:overall-system} 
Overview of a mixnet incorporating reliability estimation. Steps A1–A6 illustrate standard packet encoding and routing. Steps B1–B7 depict the additional protocol steps used for reliability estimation.
%Steps A1-A6 correspond to relaying packets during an epoch while steps B1-B7 correspond to (post-epoch) reliability estimation.
}
\end{figure*}

\begin{definition}
A mixnet with reliability estimation (REst-mixnet) is a mixnet scheme over a graph $G$ such that: (1) a publicly known number of legitimate packets $\sum_c{s}_{(c,g)}$ are submitted via each gateway $g$ during a time epoch; and (2) 
%A mixnet with reliability estimation (REst-mixnet) is a mixnet scheme over a graph $G$ where (1) time is divided in epochs; (2) in each epoch a publicly known number of legitimate packets are introduced through each gateway; and (3) 
for any adversary $\mathcal{A}$ conforming to the threat model in Section~\ref{sec:threat}, all participants can estimate the \textit{reliability} ${\rho}_e$ of a link $e=(i,j)$ in $G$, defined as the fraction of legitimate packets successfully transmitted from $i$ to $j$, relative to the total legitimate packets received by $i$ to be forwarded to $j$, i.e.: 
%i.e., including both the ${s}_e$ packets that were successfully received by $j$ and the ${d}_e$ packets that were dropped before being registered by $j$: 
\begin{equation}
    {\rho}_e = \frac{{s}_e}{{s}_e + {d}_e}
\end{equation}

Anyone with access to the broadcast channel can compute a pair  $(\hat{\rho}_e,  \epsilon_e)$ such that 
$|\hat{\rho}_e - \rho_e| \leq \epsilon_e$  holds with high confidence.% $Z$. 
%\aggelos{Add the error output for a given $Z$ confidence}
\end{definition}

%In other words, all parties are able to estimate the rate of successfully transmitted legitimate packets in individual links and even along multi-link paths of interest in the mixnet graph. 
\noindent We additionally require the following two security properties. 

\noindent {\bf Replay Protection.}
Each legitimate packet must be processed at most \textit{once} by any mix node. This is a standard property of decryption mixnets~\cite{danezis2003mixminion, piotrowska2017loopix}, where packets leave a fingerprint at each hop that can be recognized in case of replay. Duplicate packets are thus detected and dropped by the receiving node.  

\noindent {\bf Privacy.} 
The primary goal of a mixnet is to provide unlinkability between input and output packets, assuming an adversary that knows the sender and receiver sets, observes traffic in all the network links (including between clients and gateways) and even controls some of the participants. 
The reliability estimation mechanism must not leak any \textit{additional} information that could help deanonymize packets.

\section{Reliability Estimation Protocols}
\label{sec:sampling-protocol}

The core concept behind the system design is as follows: as illustrated in Fig.~\ref{fig:overall-system}, 
our reliability estimation scheme relies on users with valid credentials sending indistinguishable measurement packets through the mixnet, interleaved with their normal client traffic (steps \textbf{A.1-6}). 
%
%Clients spend a credential (step \textbf{A.1}) to gain access to the mixnet via a gateway; subsequently, they prepare packets that are interspersed with measurements at a suitable rate (step \textbf{A.2}) and send them to the gateway (step \textbf{A.3}). The gateway processes packets and verifies their validity before forwarding them to the first layer of the mixnet (step \textbf{A.4}). Mixing happens in layers ($L=3$ in the figure, step \textbf{A.5}) with mixnodes in the final layer forwarding to the exit gateway. Finally the exit gateway delivers the payload to its destination, or in case of a measurement packet it records its tag (step \textbf{A.6}) so that it engages with the protocol later. 
%
The key idea is to have gateways and mix nodes commit to the public key of credentials and packets (via unique tags) they have received (steps \textbf{B.1-2}) at the end of an epoch. These commitments to public keys, along with the measurement packets, are then opened (steps \textbf{B.3-4}) so that the measurements can be verified and traced along their paths through the mixnet (steps \textbf{B.5-7}). 
This process yields a sample of transmitted and dropped packets on each link, allowing anyone to estimate link and node reliability as we describe in the upcoming sections.

To summarize the protocols: Section~\ref{link-reliability-honest} 
describes how users generate measurement traffic and how mix nodes record received packets in a semi-honest setting.
A link reliability estimation protocol, which can be executed by anyone able to read the broadcast channel, aggregates the data broadcast by gateways and mix nodes to estimate the fraction of successfully transmitted packets on each network link, along with an associated error bound for a given confidence level.
Section~\ref{sec:vrf-based-routing} then introduces \textit{VRF-based routing} --- a cryptographic tool that generates unforgeable measurement packets according to the mixnet's routing policy and mitigates attacks by malicious clients and utility-maximizing gateways. 
Finally, Section~\ref{sec:reliability-estimation} shows how link-level measurements can be used to derive reliability scores for mix nodes themselves.

\subsection{Link Reliability Estimation in a Semi-Honest Setting}
\label{link-reliability-honest}

We begin by describing our link reliability estimation protocol in a simplified setting where clients and gateways are (semi-)honest. The steps are illustrated in Fig.~\ref{fig:overall-system}. 
%Then, we show how VRF-based routing can mitigate malicious deviations by clients and gateways. <-- redundant with ending of section intro (previous paragraph)

\noindent
{\bf Generation of packets.} A client first spends (\textbf{A.1}) a credential $c$ with a gateway $g_0$, which entitles the client to route up to $S_c$ packets through the mixnet via $g_0$.
Once the gateway acknowledges the credential, the client may submit packets. 
Each time a packet is created (\textbf{A.2}), a coin is flipped, and 
with probability $p_\mathsf{lot}$ the packet header is used to generate a measurement (green packet) (\textbf{A.3}) that is sent interleaved with the client's data and any added cover traffic (yellow packets). 
As routing intermediaries, mix nodes cannot distinguish measurement packets from other types; only the intended recipient can determine a packet’s type.

Let ${s}_{(c,g)} \leq S_c$ denote the total number of packets generated from credential $c$ and routed via $g$ during the epoch. 
Note that ${s}_{(c,g)} = S_c$ when a credential's entire allowance is consumed within an epoch, which will often be the case for small $S_c$. 
Importantly, each credential $c$ is unlinkable to any other credential from the same wallet, preventing long-term profiling of client's activities. 
The number of measurement packets ${s}_{(c,g)}^*$ follows a binomial distribution ${s}_{(c,g)}^{*} \sim B({s}_{(c,g)}, p_\mathsf{lot})$, and we expect ${s}_{(c,g)}^{*} \approx p_\mathsf{lot} \cdot {s}_{(c,g)}$. 
Measurement packets are sent interleaved with the client’s other traffic, while the \textit{commitments} to measurement packets are retained by $g$ (potentially also by the client) for later disclosure (in B.3). %\claudia{Write a remark when they are  kept by \textit{client}}

\noindent
{\bf Routing of packets.} 
Gateways forward client packets to the first-layer mix node specified in the packet’s route (\textbf{A.4}). 
Mix nodes then iteratively execute the \textit{packet processing} algorithm and relay packets to their next hop (\textbf{A.5}). 
As part of replay protection, mix nodes must record a \textit{tag} for each processed packet (denoted $t$, cf. Sect.~\ref{sec:vrf-based-routing}), e.g., computed as a hash of a secret derived via Diffie-Hellman by combining the packet header and the node’s private key.
Upon receiving a packet, the node checks whether its tag is already in the list of seen tags: if the tag is new, the packet is processed and its tag added to the list; otherwise, the packet is dropped.
Each node $j$ stores, in a space-efficient manner, the $\sum_i {s}_{(i,j)}$ tags received from its predecessors $i \in P(j)$ during the epoch. 
This \textit{tag-commitment} can be implemented via a Merkle tree~\cite{merkle1988} or a Bloom filter~\cite{bloom70}, as it only needs to be binding and space-efficient. 
We represent the nodes' tag-commitments as the red filters in Fig.~\ref{fig:overall-system}. 
When adding a fresh packet tag, the node appends to it a binary flag indicating whether the packet passed integrity checks. 
If so, it is relayed further; otherwise, it is dropped. 
Mix node decryption keys may be updated per epoch to provide forward security and prevent replay attacks across epochs.
Measurement packets are addressed to a randomly selected gateway, with selection probabilities defined by the mixnet’s routing policy. 
Upon receiving a measurement packet, the gateway stores its tag in the tag-commitment and discards the (green) packet, while forwarding on data (yellow) packets as well as cover traffic to their respective destinations (\textbf{A.6}). 
%AK: this optimization is useful but not too important i am removing it for lack of space.
%Note that gateways do not need to store in the Bloom filter the tags of non-measurement packets, meaning that their Bloom filters can be smaller than those of mix nodes.% 
%%
%\claudia{alternatively: the final destination is beyond the gateway (eg, a client of that gateway, or the validators maintaining the blockchain, or some broadcast channel), such that the gateway does not know it was a measurement packet until it is opened}\aggelos{both alternatives are viable so we have to decide how to write this.}

%The packet is also dropped if it fails integrity checks, with the tag recorded with a flag indicating that the packet was malformed and thus dropped. 
%\aggelos{do we want to say anything at this point about malformed packets and recording their tag with a flag ? if yes, this is a place where we can do it.}

%To participate in the sampling protocol, a mixnet intermediary $j$ also stores the list of seen packet tags in a Bloom, which includes the $\sum_i {s}_{(i,j)}$ packet tags received from its predecessors $i \in P(j)$ during an epoch. 

%\aggelos{Add the transmit bit with each pseudorandom tag.}

\noindent
{\bf Post-epoch stage: revelation of measurements.} Once an epoch concludes, the protocol proceeds with the broadcasting steps shown in Fig.~\ref{fig:overall-system}: 

\begin{itemize}
    \item{\textbf{B.1:}} Each gateway $g$ broadcasts the number of packets ${s}_{(c,g)} \leq S_c$ sent per credential $c$ during the epoch. 
    %Nodes and gateways broadcast a commitment (\claudia{e.g., a signed hash?}) to their Bloom filters for the epoch. In the case of nodes the filter contains entries for all the received packets while in the case of gateways only received measurements are included. 
    \item{\textbf{B.2:}} All nodes (mix nodes and gateways) broadcast their tag-commitments. 
    \item{\textbf{B.3:}} %Once all the Bloom filters are committed, the gateways %(i) announce the total number of packets ${s}_{(c,g)} \leq S_c$ sent within the epoch from each credential $c$ spent with $g$, and (ii) 
    %Each gateway $g$ broadcasts the number of packets $N_{g,c}$ sent  per credential $c$ 
    Each gateway $g$ broadcasts the openings of the ${s}^*_{(c,g)} (\approx p_\mathsf{lot} \cdot {s}_{(c,g)})$ measurement packets generated during the epoch from all its client credentials.  
    \item{\textbf{B.4:}} All mix nodes and gateways broadcast the openings of their tag-commitments corresponding to the revealed measurement packets (this step is unnecessary if Bloom filters are used, since membership can be verified directly). Note only public material is broadcast, not secret key material.
    %sent %as well as the total number of packets sent <<<--- is this still needed? 
    %from each credential $c$ they have serviced in the epoch. 
\end{itemize}

Any party observing the broadcasts can verify correctness and compute a reliability score $\hat{\rho}_e$ per link $e$:

\begin{itemize}
    \item{\textbf{B.5:}} Reconstruct the $\sum_g  \sum_c {s}^*_{(c,g)}$ measurement packets generated during the epoch (including their mixnet paths, recipient gateways and per-hop tags) using the gateway openings.
    
    \item{\textbf{B.6:}} Verify the validity of the measurement tag openings provided by mix nodes and gateways. % if Merkle trees are used as tag-commitments (step not needed with Bloom filters). 
    
    \item{\textbf{B.7:}} Determine which measurement packet tags are present in which tag-commitments, yielding the sets of transmitted ${s}^*_{e}$ and dropped ${d}^*_{e}$ measurement packets for each link $e$. Link and node reliability scores can be then derived from ${s}^*_{e}$ and ${d}^*_{e}$, as explained in the following.

\end{itemize}

Given a link $e=(i,j)$, we denote by ${s}_{(i,j)}^*$ the amount of measurement packets \textit{successfully transmitted} in the link (number of  measurement packets whose (per-hop) tags appear in the tag-commitments of \textit{both} $i$ and $j$), and by ${d}_{(i,j)}^*$ the number of measurement packets \textit{dropped} in the link, i.e., those whose tags appear in the tag-commitment of $i$ but not the one of $j$.
%and \textit{all} its predecessors\footnote{Any situation with ``skipped'' openings is indicative of either failure to record all received packets by the skipping node, or collusion between the gateway and the successor where the packet ``reappears''.\claudia{can elaborate on this as discussion point}} in the packet path -- but not in the Bloom filter of $j$. 
%We consider that no packets are dropped in the link $(c,g)$ between client and gateway, since they maintain a session that allows the client to retransmit packets if needed. 
Similarly to Eq.\ref{eq:sumpackets}, for any node $j$, all the measurement packets received from its predecessors $i$ are either successfully transmitted to their next destination $k$, or dropped before reaching $k$: 

\begin{equation}
\sum_i {s}^*_{(i,j)} = \sum_k ({s}^*_{(j,k)} + {d}^*_{(j,k)})
\label{eq:node-samples}
\end{equation}

Measurement packets revealed by a gateway $g$ and addressed to first-layer node $i$ that do not appear in $i$'s tag-commitment are considered as part of ${d}^*_{(g,i)}$. Similarly, ${d}^*_{(k,g)}$ accounts for measurement packets addressed to $g$ from last layer node $k$ that are reported by $k$ but do not appear in $g$'s tag-commitment. 
Packets reported as dropped due to failed integrity checks (as indicated by the binary flag stored with each tag) are excluded from the measurement sample. Likewise, packets whose paths exhibit `holes' --- i.e., those included in a node $j$’s tag-commitment but missing from a predecessor’s --- are also discarded as measurements.

Since mix nodes cannot distinguish measurement packets from regular ones, measurement traffic experiences the same drop behavior as user traffic. 
%Taking into account that mix nodes cannot distinguish measurement from regular packets while routing,  measurement packets are transmitted and dropped by them at the same rate as any other packet. %i.e., for any edge $e \in G$, ${s}_e^* \approx p_\mathsf{lot} \cdot {s}_e$ and ${d}_e^* \approx p_\mathsf{lot} \cdot {d}_e$. 
% It follows that the values $\hat{s}, \hat{d}$ as defined above estimate correctly the expectation of the variables ${s}, {d}$. 
%This result extends to any number of malicious clients as well, due to the fact that VRF-based routing retains its routing distribution even for malicious randomness and VRF keys.
We use this to produce a \textit{reliability score} $\hat{\rho}_e$ per link $e$, given by the transmitted and dropped samples ${s}_e^*$ and ${d}_e^*$, defined as:

\begin{equation}
\label{eq:rho_e_hat}
    \hat{\rho}_e = \frac{{s}_e^*}{{s}_e^* + {d}_e^*}
\end{equation}

The accuracy of $\hat{\rho}_e$ as an estimator of ${\rho}_e$ depends on the available number of measurement samples ${s}_e^* + {d}_e^*$, which determine the maximum sampling error $\epsilon_e$ with a given a confidence level, such that:  

\begin{equation}
\label{eq:estimation}
    |{\rho}_e - \hat{\rho}_e| \leq \epsilon_e 
\end{equation}

To bound the estimation error $\epsilon_e$ we use methods that estimate coin bias from a sequence of observed coin flips, considering that the coin flip can result in transmission (heads) with probability ${\rho}_e$; or in a drop (tails) with probability $1-{\rho}_e$. 
Without having any prior for the distribution  ${\rho}_e$, i.e., assuming all values are equally likely, ${\rho}_e$ follows a Beta distribution with density function
${\rho}_e^{s^*_e} (1- {\rho}_e)^{d^*_e} / B( s^*_e +1, d^*_e+1)$, mean
$\hat{\rho}_e$ and variance $\hat{\rho}_e \cdot (1-\hat{\rho}_e) / (s^*_e + d^*_e + 1)$.

When the number of samples is sufficiently large and the coin bias is not heavily skewed, we can apply a normal approximation\footnote{Note that this error estimation becomes increasingly inaccurate as $\rho_e$ approaches zero or one. An accurate confidence interval can be derived following Clopper Pearson \cite{clopper-pearson} (but it has no nice closed formulation as Eq.~\ref{eq:epsilon}).} that yields the {\em Wald interval}
for  the maximum estimation error $\epsilon_e$ 
at confidence level $Z$: 

\begin{equation} \label{eq:epsilon}
    \epsilon_e =  Z \cdot \sqrt{\frac{\hat{\rho}_e \cdot (1 - \hat{\rho}_e)}{{s}_e^* + {d}_e^*}}
\end{equation}

%\aggelos{need to doublecheck the above; ${\rho}_e$ is unknown and we are estimating it}\claudia{@aggelos: I added explanations, please check below}

%The maximum estimation error occurs when ${\rho}_e = 0.5$, and thus $\epsilon_e \leq \frac{Z}{2\sqrt{{s}_e^* + {d}_e^*}}$ for any value of ${\rho}_e$, with the error diminishing significantly when ${\rho}_e$ approaches zero or one. Since ${\rho}_e$ is not available, in practice we use $\hat{\rho}_e$ when computing $\epsilon_e$ as best available approximation of ${\rho}_e$. In cases where $\hat{\rho}_e=1$ or $\hat{\rho}_e=0$, we use Laplace's \textit{rule of succession} to estimate $\epsilon_e \approx \frac{1}{{s}_e^* + {d}_e^* + 2}$. 

Note that the estimation accuracy for an edge $e$ depends only on the total number of measurement packets successfully collected on it, i.e., ${s}_e^* + {d}_e^*$. 
This implies that the error bound $\epsilon_e$ is independent of the total volume of user traffic (${s}_e + {d}_e$) or the packet size. As a result, the protocol remains efficient and scalable even as the overall traffic grows.
In particular, to achieve a target error $\epsilon$, it suffices to maintain a \textit{constant} number of measurements per link, regardless of how much total traffic the link handles. 
Consequently, as traffic increases, the measurement sampling rate $p_\mathsf{lot}$ can be decreased while still maintaining estimation accuracy. 
Conversely, a sudden drop in traffic volume may yield too few measurement samples, leading to less accurate reliability estimates for that epoch; however, $p_\mathsf{lot}$ can be adjusted dynamically each epoch based on observed traffic to ensure sufficient samples for the desired accuracy. 
The output of this protocol is a value $\hat{\rho}_e \approx {\rho}_e$ for each edge e, estimating the link reliability (i.e., the fraction of successfully transmitted packets), along with a sampling error $\epsilon_e$ that quantifies the estimate’s accuracy at a given confidence level.

\noindent {\bf Implementing Tag Commitments.}
There are various options for implementing tag-commitments, including Merkle trees~\cite{merkle1988} and Bloom filters~\cite{bloom70}. Broadly speaking, the relevant cryptographic primitive  is that of a vector-commitment~\cite{DBLP:conf/pkc/CatalanoF13}. In the remaining of the paper we focus on the case of Merkle Tree based vector commitments. Beyond avoiding false positives, Merkle Trees prevent a membership-test attack feasible with Bloom filters: a client whose keys are compromised could otherwise have past packet routes reconstructed from the published filters. Moreover, Merkle Trees scale better at high traffic volume and low $p_\mathsf{lot}$.

%%%%%%%%%%%%%%%%%%%

\noindent {\bf Communication complexity.} For $n$ measurement packets, the overhead consists of transmitting these $n$ packets through the mixnet and performing $O(n \log n)$ broadcasts when Merkle trees are used for tag-commitments.
In Appendix~\ref{sec:eval-overhead} we estimate the overhead incurred in a practical deployment. % when using Bloom filters instead of Merkle trees for tag-commitments. 

\noindent {\bf Privacy.} %\claudia{to do, depends on implications of counter and info revealed at end of epoch}
%
%\claudia{not sure this belongs here or should be moved to some other section}
The scheme does not reveal any information about clients' packet routes or destinations, given that revealed measurement packets contain no client payloads and are addressed to random destinations. 
Only the total number of packets generated per credential is required, and this information is already observable by a network adversary monitoring the connection between a client and its gateway. 
Credentials are unlinkable to each other, disaggregating clients' activities over time. 
Note the tradeoff between credential size (in terms of number of packets $S_c$) and privacy: a smaller $S_c$ 
provides finer-grained disaggregation of usage per client but incurs greater overhead for credential issuance and verification, as more credentials must be handled for the same packet volume.
Ideally, credentials should have a size $S_c$ that is typically consumed within a single epoch.

\subsection{VRF-Based Routing for Link Reliability Estimation in an Adversarial Setting}
\label{sec:vrf-based-routing}
  
We now extend the link reliability estimation protocol to an adversarial setting, where clients may be malicious and gateways act as utility-maximizing entities.  A utility-maximizing gateway avoids strategies that would decrease its utility, but may otherwise deviate arbitrarily from the protocol. 
We assume a gateway utility function\footnote{As a real-world sanity check, Nym's reward function~\cite{Diaz2022Reward-mixnet} can easily be configured for gateways to increase monotonically with all these variables.} that increases monotonically with the following factors: 
(i) link reliability, as estimated by the protocol, for mixnet links involving the gateway; 
(ii) client satisfaction, defined as the fraction of a client’s submitted packets that are successfully delivered; 
(iii) credential consumption, i.e., the number of announced legitimate packets sent; and (iv) adherence to protocol specifications, such that the gateway is not caught misbehaving.

Viewed in this utility-maximizing setting, the scheme of the previous section is insufficient. 
For instance, a utility-maximizing gateway could send and open specially crafted packets that partially match the sent measurements. This can be done in such a way so that it preserves the gateway’s perceived reliability on its own links (thus maintaining its own utility), while degrading the perceived reliability of downstream links, as  it appears that those measurements got dropped %further
along their path by downstream nodes. 
Also, since clients can be malicious in our model (as e.g., in a setting where a utility-maximizing gateway employs some clients as sock puppets), we cannot rely on them to generate the measurement traffic, as they may, e.g., manipulate measurement traffic to reduce the reputation of otherwise reliable nodes. %It follows that a utility maximizing gateway can foil the reliability estimation of the previous section. 
To ameliorate these issues and demonstrate how we can still measure link reliability in this setting  we put forward a method that ensures that path selection cannot be biased: % without this property, either clients or gateways could over-generate or omit measurements to frame nodes.

\noindent
\textbf{VRF-based routing.}
%Traditionally in decryption mixnets, the client is considered honest and the mix nodes themselves may be potentially malicious, so that the client chooses the entire path through the network~\cite{danezis2003mixminion, miranda}. However, a malicious client can chose paths to bias measurements and attack nodes to overload them or lower their reputation. 
%VRF-based routing removes client control over route selection, ensuring packet paths are sampled according to the mixnet’s public routing policy.
At a high level, VRF-based routing eliminates any client or gateway control over the random values that determine packet routing parameters. Clients can only choose the payloads and final destinations for non-measurement packets. Everything else, including the
randomness used for encryption, is derived deterministically and verifiably from publicly known values. 
This enforces non-manipulable path selection and, specifically in our setting, guarantees that measurement packets cannot be forged, omitted, or selectively dropped without detection.
In more detail, our construction relies on:

\noindent
(1) A VRF function that consists of three algorithms
$\langle \vrfK, \vrfE, \vrfV \rangle$: $\vrfK$ generates a key pair $(sk,vk)$; 
$\vrfE(sk, m) = (r, \pi)$ evaluates the VRF on $m$ to produce the pseudorandom output $r\in\{0,1\}^\kappa$ as well as a proof $\pi$;
and $\vrfV(vk, m, r, \pi)$ verifies that $r$ is the correct output corresponding to 
the message $m$. 

\noindent
(2) A mixnet packet encoding scheme (e.g., Sphinx~\cite{sphinx-2009}) that sets the individual hops in a packet's path and enables its processing. 

\noindent
(3) Non-interactive zero-knowledge proofs for discrete logarithm relations, for which we use standard notation, e.g., $\mathsf{NIZK}(
(  g,y )\,\exists x :g^x = y)$ denotes a proof where the statement is a concise  description of a multiplicative group generated by $g$ as well as an element $y$ in that group and the witness is 
the discrete logarithm of $y$ base $g$. 

\noindent  
(4) A public algorithm $\mathsf{Routing}(i, r) = j$  that
for a given mixnet node $i$ and randomness $r$,
samples the node that should be the next hop  $j$, 
assuming $r$ is uniformly distributed over $\{0,1\}^\kappa$. 
A common choice for
$\mathsf{Routing}$ is to sample 
uniformly at random from the {\em successor} nodes $j \in S(i)$ in the graph $G$.  
%We assume that all node identities are bitstrings in $\{0,1\}^\ell$, where $\ell\in \mathbb{N}$ is a suitable identity length.
%
We will further require that for any $(i,j)$, the distribution of $r$ conditioned on 
$\mathsf{Routing}(i, r) = j$ is efficiently samplable. 

\noindent 
(5) A public broadcast channel, like a blockchain, that also facilitates a beacon in the form of a random nonce $\mathsf{n}$ (this value can be derived by the blockchain itself~cf.~\cite{kate2023flexirand}). 

We describe the concept of VRF-based routing in terms of encoding and processing packets. 
%To keep the presentation simple, in the main body we only provide an overview of the key ideas; for full details, we refer the reader to Appendix~\ref{app:vrf-based}. 
%
%To keep the presentation simple, we start with an overview and present a more detailed picture in section~\ref{sec:vrf-based-routing-specifics}. 
%
An overview of the notations used in the description is presented in Table~\ref{tab:notation}.

\begin{table}
\begin{center}
\begin{tabular}{|p{0.8\columnwidth}|}
\hline
$\kappa$: security parameter; length of the hash function output

$p_\mathsf{lot}$: probability of a measurement packet

$ctr$: packet counter used as VRF input

$L$: number of layers in the mixnet

$W$: width of the mixnet 

$N=LW$: number of mixnodes

$i,j,k,\ldots$: nodes in the network graph

$S_c$: number of packets enabled by credential $c$

$\rho$: reliability coefficient 

$\hat\rho$: estimation of reliability coefficient

$e=(i,j)$: link (edge) of network graph between nodes $i,j$

$s_e$: number of successfully transmitted legitimate packets along network link $e$

$d_e$: number of dropped legitimate packets along network link $e$

$s^*_e$: number of measurement packets successfully transmitted along network link $e$

$d^*_e$: number of measurement packets dropped along network link $e$

$vk$ : verification key of the VRF 

$sk$ : secret key of the VRF 

$r_\mathsf{type}$: VRF value for measurement lottery

$r_\mathsf{pkt}$: VRF value for packet encoding

$r_\mathsf{exit}$: VRF value for choosing the exit gateway
 
$y_{n}$: public-key of node $n$ 

$T$: target that determines measurement lottery outcome

$s_i$: Diffie Hellman key corresponding to packet hop $i$

$b_i$: blinding value corresponding to packet hop $i$

$\tilde{r}_i$: encoding randomness for packet hop $i$

$n_i$: node corresponding to packet hop $i$

$\mathsf{n}$:  unpredictable nonce drawn used as VRF input

$\alpha$: client chosen packet randomization key 

$\alpha_0$: packet specific randomization key

$\alpha_i$:  base element for $i$-th packet hop, for $i>0$

$\psi_i$: ciphertext for packet hop $i$ 

$x_i$: secret-key of processing node in packet hop $i$

$t_i$: pseudorandom tag calculated in packet hop $i$\\
\hline
\end{tabular}
\end{center}
\caption{\label{tab:notation} 
Notation Table. }
\end{table}

\noindent \textbf{Packet encoding.}
Packet encoding involves a client with credential $c$ and its entry gateway $g_0$. 
For a given $c$, the client selects a packet randomization key $(\alpha = g^{x},x)$ 
%\claudia{@aggelos: here it would be good to start already with the connection to the credential $c$, so that everything is in context with what came before} 
%AK: rearranged things to highlight the connection
and the gateway a VRF key pair $(vk, sk)$. 
These values are publicly committed on the public broadcast channel. 
Furthermore, both client and gateway maintain
a packet counter.  
For each packet transmission of a data packet or measurement packet by the client, three VRF outputs are computed, 
$r_\mathsf{pkt} , r_\mathsf{type} , r_\mathsf{exit} $ 
each taking as input the packet counter and a public random nonce $\mathsf{n}$ that must be unpredictable at the time when
public keys are committed. 
The key idea is to use the VRF to ``derandomize'' the mixnet encoding
ensuring two properties simultaneously. First, based on the VRF output,  
with probability $p_\mathsf{lot}$ the packet encoding is fully determined
by the VRF and corresponds to a \textit{measurement packet} ($p_\mathsf{lot}$  is the probability of the event $r_\mathsf{type}<T$, where $T$ is a suitable public target value). This ensures that measurement packets are fully reproducible by the gateway  $g_0$, who is able to open measurement tag-commitments on behalf of the client---thus removing the need for clients to be online and participate in the post-epoch steps of the protocol. 
This exploits the fact that once selection is unbiasable, who generates the measurements becomes a practical choice. Having the gateway hold and submit the measurement openings post-epoch is more practical, since gateways are typically online while clients (e.g., on mobile devices) may have intermittent connectivity.
%
%
%claudia: i'll move this point to conclusions as well
%Given that gateways are typically online while clients such as mobile devices could have intermittent connectivity, having gateways submit measurements is a more practical solution. 
%In terms of measurement generation, both the client and the gateway are able to do it. 
%\claudia{i think the longer comment about who generates those measurements is better saved for conclusions, here it feels too much like a digression}
%\claudia{@aggelos: should we include a note here saying that alternative designs may keep the openings from the gateway (only known to the client), but that they require the client to be online to submit the openings at the end of the epoch? }
%AK: we could, but i am worried it may create further confusion?
Second, non-measurement packet headers are deterministically computed based on 
both the VRF and the packet randomization
key $\alpha$, 
so that, in the view of the  gateways they remain computationally indistinguishable from regular packets under the underlying encoding scheme, hiding both their payload as well as the mixnet path they follow. 
%\claudia{@aggelos: do we need to further emphasize here that the gateway does NOT learn the packet paths for non-measurement packets?}
%AK: done

\begin{figure*}
\centering
    \includegraphics[width=0.90\textwidth]{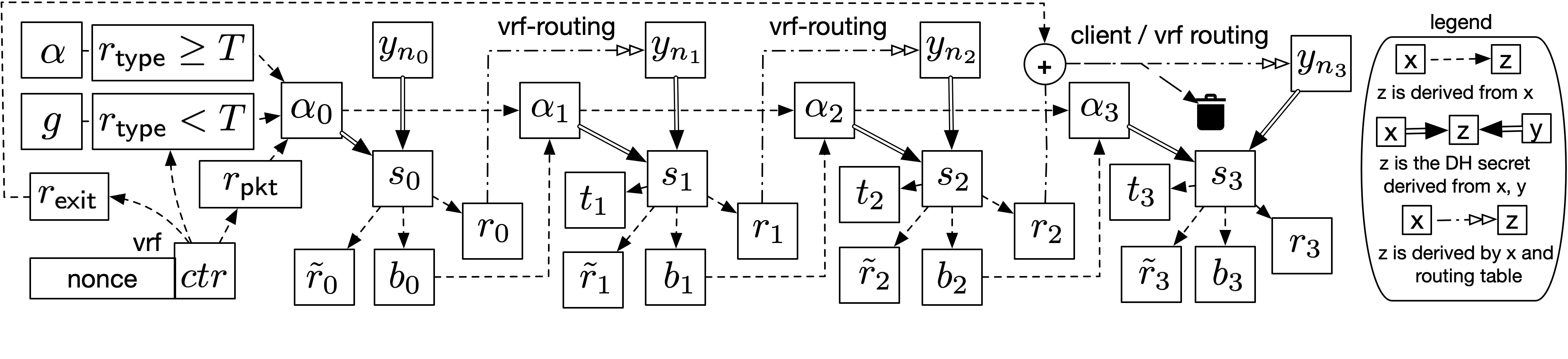}
\caption{\label{fig:dependencies-2gen} Illustrating the dependencies between the packet variables $\alpha, b, s, r,  t, \tilde{r}$  when processing a VRF-based routing packet in the case of $\nu=2$. The measurement packet condition is illustrated in the upper left.  }
\end{figure*}

To achieve these two properties, each packet involves
the calculation of a sequence of randomizer values $\alpha_0, \alpha_1, \ldots $, 
one for each hop in the packet route, where $\alpha_0 = g^{r_\mathsf{pkt}}$ or
$\alpha_0 = \alpha^{r_\mathsf{pkt}}$ depending on whether it is a measurement
or non-measurement packet.
The remaining values in the sequence 
are derived by repeatedly blinding $\alpha_0$ with pseudorandom exponents, as detailed below.
This sequence of values $\alpha_0, \alpha_1, \ldots$ is used to generate ephemeral Diffie–Hellman secrets with the public key of each node in the path, starting from the gateway $g_0$. This secret, denoted by $s_i$ at hop $i$, 
is deterministically derived from the VRF output and the randomization key $\alpha$, but appears pseudorandom to any third party. 
These secrets are used to derive
the blinding exponent for the next element in the randomizer sequence as well as to seed the mixnet encoding.
Moreover, most importantly, they also determine the next hop in the routing sequence by 
applying the $\mathsf{Routing}(\cdot ,\cdot )$ function to the identity of the 
current node and a pseudorandom value derived by the key itself (see Fig.~\ref{fig:dependencies-2gen} and Table~\ref{tab:notation}). %in Appendix~\ref{app:security- theorem}). 

A final subtlety arises when selecting the exit gateway $g_1$. 
For non-measurement packets, this gateway should be chosen at the discretion of the client;  for measurement packets, it is pseudorandomly determined by the VRF.\footnote{Note that any desired exit gateway distribution may be used here. } 
To prevent the final mix node from detecting which case applies, a value $\tilde{r}_\mathsf{exit}$ is embedded in the payload available to the last hop and xor-ed into the randomness used in the $\mathsf{Routing}$ calculation.
For measurement packets, $\tilde{r}_\mathsf{exit}$  is generated by the VRF, while for non-measurement packets is selected so that the output of $\mathsf{Routing}(\cdot ,\cdot )$  matches the desired exit gateway. 

In more details, 
suppose the client has selected an entry gateway $n_0$ with key $y_0$
    and an exit gateway $n_\nu$,
    where $\nu>0$ is the number of packet hops. 
Recall, the gateway has already established  a
VRF key $(vk, sk)$, and the client a (packet) randomization 
key $(\alpha = g^{x}, x)$. 
The gateway computes three VRF values $r_\mathrm{pkt} = \vrfE(sk,  \mathsf{n} || ctr || 
\mathsf{pkt})$, 
$r_\mathrm{exit} = \vrfE(sk,  \mathsf{n} || ctr || 
\mathsf{exit})$, 
 $r_\mathsf{type} = \vrfE(sk,  \mathsf{n} || ctr || 
\mathsf{type})$,
where $\mathsf{pkt}, \mathsf{type}, \mathsf{exit}$
are the numerical labels $\{0,1,2\}$; 
note we use $||$ to denote concatenation.
In case $r_\mathsf{type} < T$, then the packet associated with $ctr$ is a {\em measurement packet}, otherwise
we have a {\em regular packet}. 
In the latter case, 
the client receives the VRF values from the gateway,
and calculates 
$\alpha_{0} = \alpha^{r_\mathsf{pkt}}$, 
$s_0 = y_0^{r_\mathsf{pkt} \cdot x}$,
$\tilde{r}_0 = H(\mathsf{rnd}, s_0)$, 
$r_0 = H(\mathsf{next}, s_0)$,
$b_0 = H(\mathsf{bli},s_0)$, 
where $\mathsf{rnd}, \mathsf{next}, \mathsf{bli}$ are the numerical labels $\{3,4,5\}$ respectively. 
Subsequently  (for $i=1,\ldots,\nu$), 
$\alpha_i = \alpha_{i-1}^{b_{i-1}}$, 
$s_i = y_i^{r_\mathsf{pkt} \cdot x}$, where
$y_i$ is the key of node $n_i = \mathsf{Routing}(n_{i-1}, r_{i-1})$ (for $i<\nu$),
and finally, as before, 
$\tilde{r}_i = H(\mathsf{rnd}, s_i)$, 
$r_i = H(\mathsf{next}, s_i)$,
$b_i = H(\mathsf{bli},s_i)$.
In the above, we have the exception in the calculation 
$n_\nu$ so 
that 
$n_\nu = \mathsf{Routing}(n_{\nu-1}, r_{\nu-1}\oplus \tilde{r}_\mathsf{exit})$,
with 
$y_\nu$ being the key of the exit gateway and 
$\tilde{r}_\mathsf{exit}$ selected suitably so the routing to 
$n_\nu$ holds (the value $r_\mathsf{exit}$ is not used). 
In the case of a  measurement packet, 
the procedure is the same with the  distinction that $x=1$ in the calculations provided above, i.e., the randomness key $\alpha$ is not involved and as a result $\alpha_0 = g^{r_\mathsf{pkt}}$ and the gateway can calculate the measurement packet directly.
Moreover, the exit gateway is defined randomly by choosing 
$\tilde{r}_\mathsf{exit} =r_\mathsf{exit}$. 
In either case, the above process  enables to apply the mixnet encoding along the route $\langle n_0, \ldots,  n_{\nu-1}, n_\nu\rangle$, using randomness $\tilde{r}_i$ at the $i$-th hop. The resulting packet is augmented with the value $\alpha_0$.  
Note that the packet counter $ctr$ is incremented with each packet.

%For non-measurement packets this value is chosen by the client to direct the packet to its intended destination. For measurement packets, it is derived from the VRF and cannot  be influenced by either the client or the entry gateway. <-- redundant

\noindent  \textbf{Packet Processing.}
All the packets have the form $(\alpha_i, \psi_i)$, where 
$\alpha_i$ is the randomizer value associated with the packet and $\psi_i$ is the packet itself.
Assuming the processing node has public-key $y_i = g^{x_i}$, the first step
is to compute $s_i=\alpha^{x_i}$, a Diffie-Hellman secret used to process $\psi_i$.
Of particular importance is the tag value $t_i$, derived from $s_i$. This tag is stored in the tag-commitment along with a binary flag 
indicating whether all packet validation checks passed.
Some additional care is needed when the processing node is the entry  gateway $g_0$ 
or the last node of the mixnet. In the former case, recall $g_0$
controls the VRF key and can thus determine whether a packet 
is a measurement packet. In this case, the packet may be fully verified
(its computation may even be delegated by the client to the gateway). 
For a non-measurement packet, observe that $g_0$ can still verify that the $\alpha_0$ value is correctly
derived from the client’s randomization key $\alpha$ and the VRF output $r_\mathsf{pkt}$.
When the last node of the mixnet processes the packet, it extracts the value $\tilde{r}_\mathsf{exit}$ and uses
it to compute the next hop via the 
$\mathsf{Routing}()$ function. 

In more details, 
the processing of a packet $(\alpha_0, \psi_0)$ corresponding 
to VRF values $r_\mathsf{pkt}, r_\mathsf{type}, r_\mathsf{exit}$ proceeds as follows. 
The gateway calculates $s_0 =  \alpha_0^{x_0}, \tilde{r}_0 = H(\mathsf{rnd}, s_0)$, where $x_0$ is its secret-key and it verifies that: 
(I)
the mixnet encoding 
was correctly applied to produce $\psi_0$
 given randomness $\tilde{r}_0$ and the public key $y_0$; and 
(II) the per packet randomization key is correctly computed: in the case of a non-measurement packet it is $\alpha_0 = \alpha^{r_\mathsf{pkt}}$, while it is 
$\alpha_0 = g^{r_\mathsf{pkt}}$ in the case of a measurement packet. 
If these checks pass, it calculates
the next hop as 
 $n_1 = \mathsf{Routing}(n_{0}, r_0)$, where 
$r_0 = H(\mathsf{next}, s_0)$
and then
it  forwards $(\alpha_1, \psi_1)$ to node $n_1$, where $\alpha_1 = \alpha_0^{b_0}$ 
with $b_0 = H(\mathsf{bli}, s_0)$
and $\psi_1$ is the payload of packet $\psi_0$ after processing it in the way the mixnet decoding requires. 

Next we illustrate how processing works for the remaining nodes in the path. 
Node $n_i$, given $(\alpha_i, \psi_i)$ for $i=1,\ldots,\nu$,  operates as follows. 
First,  it processes the packet to reveal payload $\psi_{i+1}$ and if $i<\nu$, the next hop $n_{i+1}$.
Then it
calculates 
$s_i  =  \alpha_i^{x_i} $,
$\tilde{r}_i = H( \mathsf{rnd} , s_i) $,
and if $i<\nu$, 
$\alpha_{i+1} = \alpha_i^{b_{i}}$ 
where $b_{i} = H(\mathsf{bli}, s_i)$.
The node then verifies that (I)
the mixnet encoding 
was correctly applied to produce $\psi_i$
given randomness $\rho_i$, 
(II) in case $i<\nu-1$,
$n_{i+1} = \mathsf{Routing}(n_{i}, r_{i})$ where 
 $r_{i} = H(\mathsf{next}, s_{i})$, 
 (the randomized routing check). 
It also calculates the tag
 $t_i = H(\mathsf{tag}, s_i)$, and stores $(b_i, \mathrm{flag}_i)$ in a database of processed tags where $\mathrm{flag}_i$ is $1$ if and only if the integrity check of $\psi_i$ is valid. 
The case $i=\nu -1$ is similar but also involves pulling $\tilde{r}_\mathsf{exit}$ from the payload 
and calculating $n_{\nu } = \mathsf{Routing} (n_{\nu-1} , r_{\nu-1} \oplus \tilde{r}_\mathsf{exit})$. 
In any case, when $i<\nu$,  the node  forwards $(\alpha_{i+1}, \psi_{i+1})$ to node $n_{i+1}$. Finally, in
 case $i=\nu$,
 the payload $\psi_{\nu+1}$ is parsed as the plaintext
and is processed by node $n_\nu$.

Recall the key point of the above construction is that the randomness  used for encoding is pseudorandomly determined by the key $\alpha$ and the VRF function. Specifically, for each $ctr$, there is a per packet randomization key $\alpha_0 = g^{x\cdot  r_\mathsf{pkt}}$ that is used
to seed the mixnet encoding. 
For measurement packets,    the gateway can completely validate their computation since $r_\mathsf{pkt}$ is known to it and the secret randomization key $x$ of the client is not used in the packet calculations. The same holds true for any party that obtains the value $r_\mathsf{pkt}$.
Furthermore, any packet can be demonstrated to be a measurement or not, by checking the inequality 
$r_\mathsf{type}<T$ which is true iff  the $ctr$ value corresponds to a  measurement packet.

The following theorem is a  straightforward  consequence of the  above observations. The first and third statements follow from the Diffie Hellman assumption and the security of the VRF, the fourth relies on the VRF, while the second takes advantage of the random oracle modeling the hash function and the fact that routing choices depend on the unpredictable value of a nonce $\mathsf{n}$. A proof is provided in Appendix \ref{app:security-theorem}.

\begin{theorem}
\label{thm:vrf}
Assuming the 
pseudorandomness of the VRF, the Decisional Diffie Hellman assumption,  
the hash functions are modeled as random oracles 
and  the underlying encoding satisfies security and privacy, it holds that:
(I) given a packet $\langle \alpha, \psi\rangle$ which is routed to an honest node 
so that it 
encodes one of two payloads $M_0, M_1$, it is infeasible for the adversary
to guess which payload it contains with probability better than random
--- this is true even in the setting $vk$ is adversarially chosen;
(II) given two packets $\langle \alpha_0, \psi_0\rangle$ and 
 $\langle \alpha_1, \psi_1\rangle$ that when 
routed to an honest node   result to the processed packets 
$\langle \alpha^*_0, \psi^*_0\rangle$ and $\langle \alpha^*_1, \psi^*_1\rangle$, 
it is infeasible for the adversary to guess $b$ with probability better than random, given 
$( \langle \alpha_{b}^*, \psi_{b}^*\rangle, \langle \alpha_{1-b}^*, \psi_{1-b}^*\rangle$ 
--- this is true even in the setting $vk$ is adversarially chosen;
(III) the packet routes generated by the above process are sampled according to $\mathsf{Routing}()$ pseudorandomly, even in the setting where $vk, \alpha$ are adversarially chosen; (IV) %assuming the client is honest, 
measurement packets are computationally indistinguishable from non-measurement packets in the perspective of any mix node.
\end{theorem}

\noindent {\bf Link reliability estimation with VRF-based routing.}
We now revisit the construction of Section~\ref{link-reliability-honest} 
and augment it with VRF-based routing. We highlight only the changes. 

\noindent 
\emph{Cryptographic primitives.} 
%(I) for   e-cash scheme~\cite{DBLP:journals/popets/RialP23}, 
%(II) a discrete-log based VRF function such as ECVRF~\cite{rfc9381}.
%AK: no need for this
%where   $\vrfK$ produces a secret key $x$ and public-key $h = g^{sk_\mathsf{vrf}}$,  $\vrfE(h,x,m) = (\rho, \pi)$  where $\rho = H(m,h, u)$, $u = H(m)^{sk_\mathsf{vrf}} $ and  $\pi = (u, \pi')$ with $\pi' = \mathrm{EQDL}( g, h, u, m, \exists {sk_\mathsf{vrf}} :  \log_{H(m)} (u) = \log_g (h))$. The verification algorithm  $ \vrfV $ verifies $\pi' $ as well as $\rho = H(m,h, u)$. 
% (II) a Pedersen commitment $g^r a^m$ for public reference string $(g,a)$,
The VRF function can based on Dodis Yampolskiy's PRF \cite{DBLP:conf/pkc/DodisY05}; specifically,    $\vrfK$ produces a secret key $sk_\mathsf{vrf}$ and public-key $vk_\mathsf{vrf} = g^{sk_\mathsf{vrf}}$,  while the VRF evaluation is set to $\vrfE(\mathsf{n}|| ctr || \mathrm{label} ) = (r_\mathsf{label} , \pi)$  where $r_\mathsf{label} = H(vk_\mathsf{vrf}, g,  u_\mathrm{label})$, $u_\mathrm{label} = g^{1/(sk_\mathsf{vrf} + 2^\ell \mathsf{n} + 4 \cdot ctr + \mathrm{label} + 1 )} $, 
 and  $\pi_\mathrm{label} = (u_\mathrm{label},  \pi'_\mathrm{label})$ 
 % PCalc: base element + proof of quality of discrete log
with
 $\pi'_\mathrm{label} = \mathrm{NIZK}( g, vk_\mathsf{vrf}, u_\mathrm{label}, ctr, \ell, \mathsf{n}, \allowbreak  \mathrm{label}:  \exists sk_\mathsf{vrf}(   
 u_\mathrm{label}^{sk_\mathsf{vrf}+2^\ell \mathsf{n} + 4 ctr + \mathrm{label}+1})$ $ || vk$ : verification key of the VRF  $g  \land vk_\mathsf{vrf} = g^{sk_\mathsf{vrf}}))$.
%
% PCalc: proof of equality of discrete log (c,s1, s2) = 1 hash + 2 exponents 
%
 Note in the above  $\ell, \mathrm{label}\in\mathbb{N},$ are public values and $ ctr< 2^{\ell-2}$ always where $ctr$ is the packet counter.

\noindent 
\emph{Generation of packets.} Step \textbf{A.1} is augmented to include the setup of the public values $\alpha$ and VRF key $vk$. 
Together with the acknowledgment of credential, the gateway shares the VRF secret-key $sk$ with the client. 
Subsequently, step \textbf{A.2} follows 
the packet encoding process of VRF-based routing, with probability $p_\mathsf{lot}$
determined by threshold $T$. In addition to measurement and 
regular non-measurement packets, we assume the client
also sends special non-measurement packets called \textit{self-loops}. 

\noindent \emph{Self-loops.}
A self-loop is a packet that a client sends to itself via the mixnet. 
In prior work~\cite{piotrowska2017loopix} and in the deployed Nym mixnet,\footnote{Nym clients currently send a self-loop every $200$ms (on average), as well as sending self-loops to fill gaps in a Poisson schedule (packets sent every $20$ms on average) when there are no data payloads to send.} self-loops are a type of cover traffic that enhances privacy by providing unobservability properties. 
In our context, self-loops additionally enable a client to monitor the reliability of the entry gateway it has selected, by checking whether packets are successfully delivered (the gateway cannot distinguish such cover traffic from data packets received for that client). 
A client can switch to a different gateway (with a new credential) if it finds the current gateway’s reliability unsatisfactory.

Steps \textbf{A.3} and \textbf{A.4} remain unaltered,
with the only difference being that packet integrity checks are enhanced 
by incorporating how the VRF is used to compute the randomizer values.
Steps \textbf{A.5}, as well as \textbf{B.1}-\textbf{B.2}, remain unchanged. 
Step \textbf{B.3}, however, is modified to reflect 
the use of VRFs in generating measurement packets: 

\noindent \emph{Opening a measurement packet.} Opening a measurement packet corresponding to a credential containing $\alpha$ is done 
by revealing the values $(ctr, r_\mathsf{pkt}, \pi_\mathsf{pkt},  r_\mathsf{type},   \pi_\mathsf{type} , r_\mathsf{exit}, \pi_\mathsf{exit})$. 
In this case, anyone can verify in step \textbf{B.5} that the packet is a measurement packet corresponding to counter $ctr$  by checking that 
$r_\mathsf{type}<T$ and validating the proofs. Moreover, the full computation of the packet is verifiable, as the packet can be deterministically regenerated from the values $r_\mathsf{pkt} $ and $r_\mathsf{exit}$, together with the packet counter $ctr$ and the public nonce $\mathsf{n}$.  
As for the size of the opening, it requires one counter value, one group element, one hash, and two exponents, for each of the three VRF openings.
Instantiating the base group via Ed25519,  %256 bits + 256 bits for the hash + 2 exponents of 256 bits + 32 bits for the counter (10 billion packets)
% 1056 bits per proof = 132 Bytes per VRF opening .
results in $388$ %396-4-4 
bytes per measurement opening.

\ignore{%%%  below does not work
\noindent {\bf Packet count validation.} 
Gateway $g$ announces in the broadcast channel the number of packets sent per credential, say $N_{g,\alpha}$;  mix-node $n$ of layer $1$ in the mix-net announces the total number of packets they received by gateway $g$, denoted by $N_{g,n}$. \claudia{note to @self: revise the following considering sum of receipts from layer 1 nodes rather than median}\aggelos{sum of receipts will not be resilient to byzantine behavior of layer 1 mix nodes } \claudia{true -- also, we may want to consider using two-party signed receipts per link rather than announcements, and do so for all links and not just gateway-layer1} Let $N_g^\mathsf{gw} = \sum_{\alpha} N_{g,\alpha}$ and $N_g^\mathsf{mix} = W\cdot \mathsf{median}_{n}( N_{g,n} )$, where $W$ is the width of the mixnet. 
Note that it should be the case that $N_g^\mathsf{mix} \approx N_g^\mathsf{gw}$; in case 
$N_g^\mathsf{mix} \not\in (1\pm \epsilon) \cdot N_g^\mathsf{gw}$ we pronounce the gateway is adversarial. 
}%%%%%%%%%%%%%%%%%%%%% 

\ignore{ %%%% Looks like we don't need those for now
\noindent {\bf Proof of counter bounds.} 
At the end of an epoch the gateway will announce the number of packets sent by each of the credentials it is serving. It follows that this will create a sequence of pairs of the form $\langle \alpha_i, \mathsf{max}_i\rangle$  for $i=1,\ldots, n$. Now recall that for each of the challenged packets in the proofs of packet legitimacy when $ch\in \{ \mathsf{left}, \mathsf{right} \}$, 
there is an opened path in the Merkle tree that reveals tha values
$\tau_j^\mathsf{cred} = \alpha(i)^t$  
and
there is a  $C_{ctr} = g^{r_\mathsf{ctr}} a^{ctr}$, commitment signed by the gateway that, assuming the gateway is honest, 
must satisfy $ctr \leq \mathsf{max}_i$. To prove this, a set of pairs $(i,j)$ is drawn based on public randomness (in our construction this will originate from the blockchain), 
and the gateway must provide a proof that $g^t = G $ and
$$  (C_\mathrm{ctr} = g^{r_\mathsf{ctr}} a^{ctr} \land \mathrm{BP}( ctr \leq \mathsf{max}_i)   \land
\tau_j^\mathsf{cred} = \alpha(i)^t) \lor (  \tau_j^\mathsf{cred} \neq \alpha(i)^t ) 
$$
where $\mathrm{BP}(ctr\leq \mathrm{max}_i)$ is a Bulletproof range proof \cite{DBLP:conf/sp/BunzBBPWM18,bulletproofsplusplus} that establishes the stated upper bound, and the notation $G=g^t \land (A \neq g^{t})$ denotes a proof of inequality of discrete logarithms, \cite{DBLP:conf/crypto/CamenischS03}.  
}%%%%%%%%%%%%%%%%%%%%%%%%%

\noindent \emph{Proof of no-skipping.}
In addition to the above, at step \textbf{B.3}, the gateway must demonstrate
that it sent all measurement packets. Specifically, for any counter value $ctr$ that does not correspond to a measurement packet, 
the gateway can open the VRF value corresponding to this counter in the packet sequence and demonstrate that it corresponds to a non-measurement legitimate packet.
Namely, the gateway reveals the tuple $\langle ctr, r_\mathsf{type}, \pi_\mathsf{type} \rangle$, which can be verified 
against $vk$ using the test $r_\mathsf{type}\geq T$. 
Given that all measurement packets have been opened, a random subset of the complement set of non-measurement counters can be challenged in this way to ensure that no measurement packets have been skipped.  
The randomness can be derived from a public nonce that is drawn after steps \textbf{B.1} and \textbf{B.2} have completed. 
In more detail, suppose we want to ensure that the amount of 
skipping is below a certain threshold, say, $\tau_\mathsf{s}$.  
Then by testing $v$ random non-measurement packets, if the skipping
rate is at least $\tau_\mathsf{s}$, we will fail to detect it
with probability $(1-\tau_\mathsf{s})^v$. 
It follows that we can set $v = \log(1/\epsilon)/\log(1-\tau_\mathsf{s})^{-1} $,
where $\epsilon$ is the desired error probability. 
In terms of size of a single opening, we must reveal one index,  one group element, one hash, and two exponents
for the proof. Using Ed25519, this amounts to 
%256 bits + 256 bits for the hash + 2 exponents of 256 bits + 32 bits for the counter (10 billion packets)
 %1056 bits per proof = 
 132 Bytes per counter position.

\ignore{%%% 
 \aggelos{XXX TO DISCUSS XXX}
\noindent {\bf Proof of routing.}  When a client delivers a non-measurement packet to a gateway, it may elect to open it later to challenge the gateway. To make this work, the gateway issues a signature to the client on each non-measurement packet as ``promise to forward.'' At the end of the epoch, the client can open the packet revealing the VRF values $r_\mathsf{type}, r_\mathsf{pkt}, r_\mathsf{exit}$ and the associated proofs. Once the packet is opened it can be traced in the Bloom filters of all nodes. As in the case of a measurement packet this opening requires \aggelos{388 bytes + the packet + the signature} assuming Ed25519. \aggelos{There is an easy framing attack against this approach, where a client and a first layer mix-node can frame the gateway.}
}%%%%%%%%%

We next illustrate how these adjustments 
to the link estimation protocol 
of Section~\ref{link-reliability-honest} 
mitigate all attacks in our threat model.  

\noindent {\bf Analysis.}
The protocol augmented with VRF-based routing is also extended by having gateways broadcast proofs of no-skipping (as a second part of step B.3 in Fig.~\ref{fig:overall-system}), ensuring that  (almost) all measurement packets are accounted for. 
The first key observation is that, conditional on a legitimate packet being sent by a gateway,  we obtain  a coin flip  with probability $p_\mathsf{lot}$ for sending a measurement packet. 
Due to gateways being utility-maximizers, observe that the total number of announced packets sent by gateway $g$,  $S_g = \sum_{c} s_{(c,g)}$, is at least as large as the actual number of packets sent. 
%\claudia{there's a mismatch between two notations for credential, which is denoted as $c$ in most of the paper but as $\alpha$ in this section}\aggelos{i fixed the above -- good catch it should be $c$.  I don't think there is an issue otherwise as $\alpha$ is just a specific base element that is part of $c$. }

We next require that entry gateways open all the measurement packets in the announced ranges for each credential. Due to the proof of no skipping, any gateway that does not want to get caught lying must open all the measurement packets in the range for each credential it announced.
Moreover, note  that if there is a credential that is misreported to have been used less than in reality, e.g., the declared $s_{(c,g)}$ is smaller than the actual $s^\mathsf{a}_{(c,g)}$, then there must exist another credential $c'$ reported by the same gateway that compensates for it,  i.e., such that $s^\mathsf{a}_{(c',g)}<s_{(c',g)}$. 
It follows that for credential $c'$ there are measurement packets that have not been opened --- thus exposing gateway $g$, who will be caught  in the proof of no-skipping for credential $c'$. %\claudia{text above is hard to understand -- clarify}\aggelos{i expanded it - note that again it is a statistical argument - we can also include the exact calculations}
Furthermore: 
(1) If measurement packets are not sent (despite being opened), this will reduce the measured reliability of the gateway, so a utility-maximizing gateway will not perform this action. 
(2) If non-measurement legitimate packets are selectively not sent (despite having the opportunity to send them) 
there are two possibilities: (i) the non-measurement packets 
are simply dropped, in which case client satisfaction decreases and thus the gateway will not prefer this strategy; (ii)   
a gateway that colludes with the client may substitute legitimate packets by illegitimate packets that carry the same payload --- this does not affect our estimation, which is only concerned with the reliability of legitimate packets. 
%\claudia{how could the gateway do this without colluding with the client?? the gateway has no access to the client payloads!} \aggelos{this statement is meant for a setting where client and gateway collaborate e.g., the client uses tweaked software that the gateway provides. - we can clarify.}

For receiving gateways, assuming  they are utility-maximizers, 
we observe: 
(1) If measurement packet tags are not recorded by the gateway in the tag-commitment despite being  received, the measured reliability will drop, hence 
the gateway will not perform this action. 
(2) If non-measurement packets 
are not handed over to their destination 
this means that they have been processed, identified
as non-measurement and then dropped. 
This affects client satisfaction (clients do not receive response traffic and are likely to switch to another gateway), and as a result, is not a preferred strategy of a utility-maximizer gateway. 
In summary, with VRF-based routing, there is no strategy available to a utility-maximizing gateway (or malicious client) that can bias link reliability estimation in the mixnet.
% and hence the results of the previous section carry to this setting. 

\ignore{%%%  looks like we don't need this
\noindent {\bf Proof of non-measurement packet.}
This proof also takes place at the end of an epoch.
At that time, 
the indices of the measurement packets have been opened. For each regular packet which was challenged with 
$ch\in \{ \mathsf{left}, \mathsf{right} \}$, the gateway 
proves that the packet was indeed a non-measurement packet. Observe that it is critical 
for privacy at this stage to hide the counter values. 
We illustrate how the proof is achieved for the case $ch = \mathsf{left}$, 
the other case is symmetric. The gateway
prepares 
the $\mathrm{NIZK}$ proof such that $\exists r_\mathsf{ctr}, ctr, sk_\mathsf{vrf}:  $
$$  C_\mathsf{ctr} = g^{r_\mathsf{ctr}} a^{ctr} \land  u_\mathsf{left}^{sk_\mathsf{vrf}+ 2^\ell \mathsf{n} + 4 ctr + \mathrm{left} +1} = g \land  \tau^\mathsf{vrf}_j = G^{sk_\mathsf{vrf}},$$
        as well as it opens the commitment  to $u_\mathsf{type}$, 
         $H(r'_\mathsf{type}, u_\mathsf{type})$, 
         so that the non-measurement condition can be verified  by checking the relation $r_\mathsf{msr} = 
          H(vk_\mathsf{vrf}, g, a, u_\mathsf{type}) >T$.
}%%%%%%%%%%%%%

\ignore{%%%% AK: not clear if the below is useful for anything
\noindent 
{\bf Validating a regular packet.} A client can also open the VRF values of a regular packet that corresponds to the $j$-th payment, revealing the values $(j, ctr, \allowbreak  r_\mathsf{pkt}, r_\mathsf{msr}, \pi_{ctr})$. In such case (I) anyone can verify that this is a regular packet corresponding to counter $ctr$ and payment $j$ by checking that 
$r_\mathsf{msr}\geq T$ and 
$\vrfV( vk_{\mathsf{vrf}},
(\mathsf{n}||r_\mathsf{pkt}||r_\mathsf{msr}), \pi_{ctr})=1 $. Note that contrary to the case of a  measurement packet, this action does not enable 
the public repetition of every step of the computation. 
}%%%%%%%%%%%%%%%%%% 

%%%% REMOVE

\subsection{Node reliability estimation}
\label{sec:reliability-estimation}

The described reliability estimation protocol provides \textit{per-link} reliability estimates. 
However, what is needed for effective network management, e.g., to determine which nodes to include in the mixnet, is a \textit{per-node} reliability score. 
Each link involves two nodes: a \textit{predecessor} who sends packets and a \textit{successor} who receives them, so estimating node reliability entails distributing responsibility for observed packet drops between the two endpoints of a faulty link. In this section, we show how to derive node reliability scores from the set of successful and failed link measurements ${s}^*_e$ and ${d}^*_e$.

The goal of this protocol is to estimate, for every node $j$ participating in the mixnet,\footnote{Nodes not actively participating in the mixnet in an epoch need to be tested separately in order to update their reliability scores---a challenge outside the scope of this work.} a reliability score $\hat{\rho}_j$ that reflects the rate of packet losses attributable to $j$, where $\hat{\rho}_j=0$ means that $j$ dropped all packets sent to it (e.g., if $j$ is offline for the entire epoch) and $\hat{\rho}_j=1$ indicates that the node was consistently online and correctly processed and forwarded all received packets. 
These $\hat{\rho}_j$ can be computed by any entity that knows the outputs of the  protocol described in Sect.~\ref{link-reliability-honest}; that is, the counts ${s}^*_e$ and ${d}^*_e$ of successfully transmitted and dropped measurement packets per link $e \in G$. 
We consider reliable inter-node transmissions such that any packet drops are either due to a failure of the link predecessor or the successor. 
%(i.e., ${\xi}_{e}=0$ for all edges).
The estimated node reliability $\hat{\rho}_j$ is computed as:

\begin{equation}
    \hat{\rho}_j = \frac{\sum_{k \in S(j)} ({s}^*_{(j,k)} + \hat{\beta}_{(j,k)} {d}^*_{(j,k)})}{\sum_{i \in P(j)}  ({s}^*_{(i,j)} + \hat{\beta}_{(i,j)} {d}^*_{(i,j)})}
\label{eq:estimated_node_perf}
\end{equation}

\noindent
In the case of gateways, $\hat{\rho}_g$ is computed considering the successfully transmitted (and dropped) measurements
to first-layer node successors $k \in S(g)$ ($g$'s outgoing links), and the received (and dropped) measurements from last-layer node predecessors $i \in P(g)$ as recorded in their tag-commitment ($g$'s incoming links). 

The parameter $\hat{\beta}_{e}$ denotes the fraction of drops \textit{attributed} to the link successor, with the remaining $1-\hat{\beta}_{e}$ being attributed to the link predecessor. 
Ideally, participants verify that $\hat{\beta}_e \approx {\beta}_{e}$, where ${\beta}_{e}$ is the \textit{actual} fraction of drops caused by the link successor. 
In a network deployment where packets are transmitted via point-to-point links $e=(i,j)$, it is however not feasible for third parties to ascertain the true value of ${\beta}_{e}$, and thus $\hat{\beta}_e$ must be estimated based on observed failure patterns and available evidence across the entire network. 
We therefore adopt a \textit{threshold} based approach to estimate $\hat{\beta}_{e}$, which considers the observable behavior of the 
two counterparties of each link across all of their connections.
%, taking into account the link reliability scores $\hat{\rho}_e$ (Eq.~\ref{eq:rho_e_hat}) across the set of incoming and outgoing edges of each node. 

Given the link reliability estimates $\hat{\rho}_e$ for $e \in G$ (computed according to Eq.~\ref{eq:rho_e_hat}), we define for each node $j$ the median incoming link reliability $\bar{\rho}^{\mathsf{in}}_j$ as the median of the set 
$\{\hat{\rho}_{(i,j)} \mid  i \in P(j) \}$, and the median outgoing link reliability $\bar{\rho}^{\mathsf{out}}_j$ as the median of the set $\{\hat{\rho}_{(j,k)} \mid  k \in S(j)\}$
These median values are then compared to a preset threshold $\bar{\tau}$. 
If node $j$'s $\bar{\rho}^{\mathrm{in}}_j \geq \bar{\tau}$ (respectively, $\bar{\rho}^{\mathrm{out}}_j \geq \bar{\tau}$), then $j$ is considered reliable on its input (respectively, output) links; otherwise, it is considered unreliable. In this way, each node receives two labels (one for inputs and one for outputs), each independently set to either \textit{reliable} or \textit{unreliable} based on the median link reliability in each direction.
Given a link $e=(i,j)$, if both $i$'s output and $j$'s input are \textit{reliable}, or both are \textit{unreliable}, then drops ${d}^*_{(i,j)}$ are symmetrically attributed to $i$ and $j$ (half to each with $\hat{\beta}_{(i,j)}=\frac{1}{2}$). 
If only one of the two is unreliable, then all the drops in that edge are attributed to the unreliable node ($\hat{\beta}_{(i,j)}=0$ if $i$'s output is \textit{unreliable} and $\hat{\beta}_{(i,j)}=1$ if $j$'s input is \textit{unreliable}). This mechanism effectively identifies unreliable nodes experiencing issues such as downtime or insufficient throughput, as these issues impact a node’s median reliability.
In adversarial settings, malicious nodes engaging in selective packet dropping also suffer degradation in their own reliability scores, proportional to the harm they cause to otherwise reliable peers.
We refer the reader to Appendix~\ref{appendix-beta} for further details on the rationale and implications of various approaches to assigning values to $\hat{\beta}_e$, and additional explanations why the threshold approach outlined above (and elaborated in Appendix~\ref{appendix-beta-threshold}) offers the best tradeoffs.  

%, even for the nodes that are part of the edge: $i$ cannot distinguish whether a sent packet was dropped by $j$ or the underlying connection; and if the $j$ did not receive a packet, it cannot distinguish whether it was never sent by $i$, or sent but dropped in the underlying connection. Even if $\overset{\triangle}{\xi}_{(i,j)}=0$ and all drops are either due to $i$ or $j$, they can blame each other for the drops and it is not possible for third parties to establish who is telling the truth. 
%In practice, it is thus not possible to compute $\overset{\triangle}{\rho}_j$ in a deployed system, though still possible to do so in a simulated environment, as we do in the next section, where the ``God's eye view'' is available. We define $\hat{\rho}_j$ as an estimation of $\overset{\triangle}{\rho}_j$, computed as:
%The policy for setting $\beta_e$ values is a design choice that affects the attribution of drops regardless of who `truly' is to blame in each specific instance. 

%Observe that an offline node $j$ would exhibit an increased drop rate acrossall incoming or outgoing links. 

\noindent
{\bf Reliability estimation in unreliable settings.} 
We first consider scenarios where all nodes are non-adversarial but up to half of each layer may be unreliable -- going offline, experiencing limited throughput, or otherwise subject to random failures -- and argue that the estimated $\hat{\rho}_j$ is a good approximation of the underlying ${\rho}_j$ (${\rho}_j$ is given by Eq.~\ref{eq:estimated_node_perf} but using `ground truth' ${s}_e$, ${d}_e$ and ${\beta}_e$ instead of ${s}^*_e$, ${d}^*_e$ and $\hat{\beta}_e$). Note that these ground-truth values are only accessible in simulation settings.
Suppose that $j$ is a reliable node that correctly receives, processes, and forwards all packets. Then, for any predecessor $i\in P(j)$, 
either ${d}_{(i,j)} = 0$ (if $i$ is also reliable), or 
${\beta}_{(i,j)}= 0$ (if $i$ is unreliable). 
With respect to successors $k\in S(j)$, either ${d}_{(j,k)} =0$ (if $k$ is reliable), or
${\beta}_{(j,k)}=1$ (if $k$ is unreliable). 
It follows that ${\rho}_{j}=1$. 
Assuming that at least half the nodes in every layer are reliable, when $j$ is reliable and communicates with a majority of reliable nodes in both the preceding and succeeding layers, 
its median values
$\bar{\rho}^{\mathsf{in}}_j,\bar{\rho}^{\mathsf{out}}_j$ are guaranteed to be above the reliability threshold $\bar{\tau}$, ensuring that $j$ is correctly labeled as \textit{reliable}. If a predecessor $i$ or a successor $k$ are unreliable, 
it holds that $\hat{\beta}_{(i,j)} = 0 $
and $\hat{\beta}_{(j,k)} = 1$. This leads to $\hat{\rho}_j \approx 1 = {\rho}_j$. 
In the case where $j$ is unreliable,
observe that for reliable predecessors $i$ and successors $k$, we also have ${\beta}_{(i,j)} = 1 = \hat{\beta}_{(i,j)}$
and ${\beta}_{(j,k)}=0 =\hat{\beta}_{(j,k)}$, thus correctly attributing packet drops to node $j$. 
On the other hand, 
if either $i$ or $k$ are also unreliable, discrepancies in downtime between them and $j$ can lead to minor deviations in the estimate  $\hat{\rho}_{j}$ relative to the true value ${\rho}_{j}$, 
since the algorithm will set the $\hat{\beta}_{(i,j)}$ or $\hat{\beta}_{(j,k)}$ to $1/2$, effectively averaging out the two unreliable performances. We 
investigate this experimentally in Sect.~\ref{sec:experiments-unreliable}
and demonstrate that $\hat{\rho}_{j}$ remains a good approximation of ${\rho}_{j}$, with accuracy improving as the number of measurement packets increases. 

\noindent
{\bf Reliability estimation in adversarial settings.} 
Adversarial nodes that are unreliable and fail to receive some packets addressed to them have no way of including the necessary evidence in their tag-commitments to falsely appear \textit{more} reliable than they truly are. 
An alternative attack strategy is to degrade the estimated reliability of a target node $j$ (or a set of targets), such that $\hat{\rho}_j < {\rho}_j$. 

We now consider an adversarial scenario in which a colluding set of malicious nodes drops packets to and from target honest nodes they wish to discredit. Rather than obtaining an immediate benefit, such adversaries may seek to damage the target’s reputation in order to create opportunities for their own nodes~\cite{dingledine2003reliable}. 
For example, in Nym, %the estimated reliability %and the amount of stake delegated to 
%of a node determines its participation and rewards.
a low estimated reliability can result in a node’s temporary exclusion from selection for the mixnet, thereby opening up space for adversarial nodes. 
%negatively affects node profitability, driving away the node's stake delegations -- which adversaries may hope to attract for their own nodes.
Our goal is to ensure that an adversary launching such an attack incurs an aggregate degradation in its own estimated reliability that is at least comparable to the aggregate degradation it inflicts on its targets. This addresses a key shortcoming of previous work, where adversaries could gain a relative reputational advantage over targets through equivalent attacks~\cite{dingledine2003reliable}. 
%If this is the case, the adversary fails to achieve its ultimate goal, since adversarial nodes will also exhibit poor reliability and will also lose, rather than increase, their rewards and  reputation with a reliability score lower than if they did not engage in any attack, %and attractiveness to staking delegates,
%thus paying a price for engaging in the attack. 
Thus, although selectively dropping packets to lower others’ reliability scores cannot be entirely prevented, it can be disincentivized by ensuring that the adversary proportionally loses utility compared to not deploying the attack. Fully preventing adversarial packet drops would require every link between entities to operate over a public channel, such as a “bulletin board” (as used in e-voting mixnets), which is incompatible with low-latency communication requirements.

Consider $j$ to be an honest and reliable node targeted by the attack. As long as malicious nodes make up less than half of the nodes in the layer preceding and succeeding $j$, the best the adversary can do is drop packets directed to $j$ in a way that results in both counterparties sharing the blame. This occurs because the adversary cannot make $j$ appear unreliable -- $j$ is connected to a majority of reliable nodes, so its median reliability values $\bar{\rho}^{\mathsf{in}}_j,\bar{\rho}^{\mathsf{out}}_j$ remain above the threshold). 
It follows that the adversary’s optimal strategy is to keep its own nodes reliable (by maintaining good reliability across most incoming and outgoing links) while triggering the algorithm to assign blame equally $\hat{\beta} = 1/2$ on any links shared with $j$. 
In Sect.~\ref{sec:experiments-adversarial} we validate this intuition empirically, demonstrating the proportional degradation in reliability observed by both adversarial and targeted nodes.

%Finally, we note for the case of the first and last layer of the mixnet, the same arguments as above apply under the assumption that the majority of legitimate traffic originates from honest gateways. Note also that gateways have an interest in routing (rather than dropping) traffic from and to their own clients, as those clients will otherwise move away to a different gateway if the quality of service is too low. 

%\section{Protocol evaluation}

\section{Empirical evaluation}
\label{sec:empirical-evaluation}

We have implemented a discrete-event mixnet simulator to 
empirically assess the accuracy of the proposed node reliability estimation approach.
Using simulation enables us to model a range of network conditions---including reliable, unreliable, and adversarial node behavior---for which the ground truth is known. This allows us to compare each node’s \textit{true} reliability score, ${\rho}_j$, against the \textit{estimated} score, $\hat{\rho}_j$, derived from the available measurement samples.
Appendix~\ref{appendix-simulator} provides a detailed description of the simulator implementation, its outputs, and the experimental setups used to evaluate the system under both unreliable and adversarial conditions.

\begin{figure}
    \includegraphics[width=.98\columnwidth]{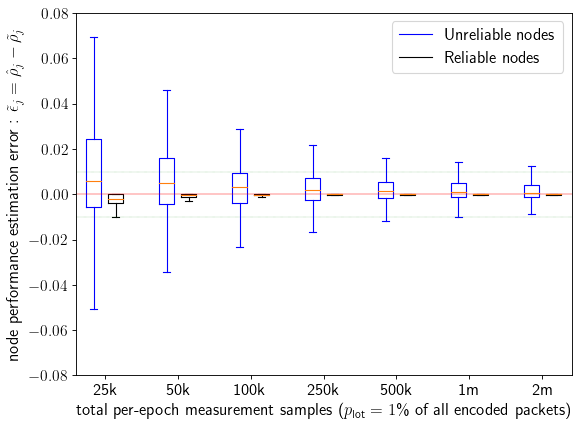}
    \caption{Distribution of error ${\epsilon}_j = \hat{\rho}_j - {\rho}_j$ for unreliable (blue) and reliable (black) nodes. %, for %different amounts of measurement samples (from 
    %$25$k to $2$ million measurements per epoch. 
    %; outliers are not depicted.
    }
    \label{fig:perf-reliability}
\end{figure}

\noindent
{\bf Unreliable setting.} 
\label{sec:experiments-unreliable}
%\claudia{thinking that a neat experiment would be to have a free riding node and see how its reliability/rewards degrade over the epochs as it gets blacklisted by more and more nodes. This requires \textbf{major} upgrades to the current simulator, so not something I can do immediately.}
We begin by evaluating node reliability estimation in a setting where all nodes are honest, but up to half the nodes in each layer may be unreliable. % as detailed in Appendix~\ref{exp-setup-reliability}.  
To assess estimation accuracy for varying amounts of measurement packets, we consider seven simulation setups with between $25,000$ and $2$ million measurement packets per epoch. These correspond to between 2.5 and 200 million total packets for $p_\mathsf{lot}=1\%$. 
Figure~\ref{fig:perf-reliability} shows the resulting distributions of the estimation error ${\epsilon}_j = \hat{\rho}_j - {\rho}_j$, distinguishing between reliable (black) and unreliable (blue) nodes.
Simulation results are aggregated over 20 runs for each setup, each resulting in $320$  values of ${\epsilon}_j$ (i.e., $6400$ samples per pair of boxplots). 
%Each boxplot shows the median (orange line), the first and third quartiles (upper and lower limits of the box), and the range of the distribution (whiskers).
Note that for reliable nodes, ${\rho}_j=1$, so their estimation error is always ${\epsilon}_j \leq 0$, since their true reliability cannot be overestimated. 
For unreliable nodes, i.e., those with ${\rho}_j < 1$, the error ${\epsilon}_j$ may be either positive or negative. 
As shown in the figure, the reliability estimation error is consistently small for reliable nodes (black boxes), becoming negligible once $100$k or more measurement samples are used.
%Thus, given enough samples, in this setting the proposed protocol provides \textit{highly accurate} reliability estimates for reliable nodes. 
The estimation error is larger for unreliable nodes but decreases with more samples: with $25$k samples the error range exceeds $7\%$ of overestimation and $5\%$ underestimation, while with $2$ million samples the error is contained within $\pm 1\%$. 
Note that nodes are rearranged in the mixnet each epoch and measured independently, so we expect per-epoch estimation errors to average out when reliability is tracked across many epochs.

\noindent
{\bf Adversarial setting.} 
%\subsubsection{ setting}
\label{sec:experiments-adversarial}
We next evaluate the proposed node reliability estimation protocol under adversarial conditions. We consider an adversary that controls a set $A$ of malicious nodes and aims to degrade the measured reliability of a set $T$ of honest and reliable target nodes.
The sizes $|A|$ and $|T|$ range (independently) from a single node up to  $40\%$ of two non-adjacent mixnet layers ($32$ out of $80$ nodes per layer). %The details of the experimental setup are provided in Appendix~\ref{appendix-setup-adversarial}. 
To attack a target node, an adversarial node must share an edge with the target---either as its predecessor or successor. If the adversary is the successor, it simply drops packets coming from the target without recording them in its tag-commitment. If the adversary is the predecessor, it registers the received packet tags in the tag-commitment, but drops the packets instead of forwarding them to the target.
The goal of the attack is to degrade the measured reliability of the targets relative to their true reliability. The more the protocol underestimates the reliability of the target nodes, the more effective the attack. The adversary’s cost is the ``opportunity cost'' incurred from its own degraded measured reliability, due to having dropped packets. The attack is thus more costly when the protocol assigns a lower reliability score to adversarial nodes than what they would have received had they behaved honestly.

%\subsubsection{reliability Penalty for Adversaries and Targets}

%\noindent
%{\bf Combining Adversarial and Unreliable nodes.}
\label{sec:perf-adversarial}

\begin{figure}    
\centering
\includegraphics[width=0.98\columnwidth]{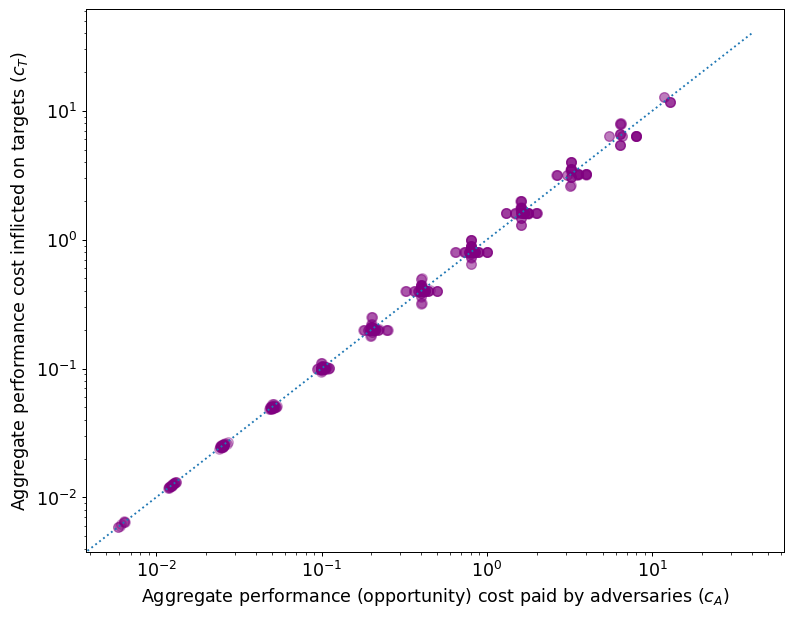}
\caption{Results for adversarial settings where all honest nodes are reliable. % Each sample represents the aggregate reliability penalties $c_T$ for $1\leq |T|\leq 64$ targets ($y$ axis) and $c_A$ for $1\leq |A|\leq 64$ adversaries ($x$ axis) in each simulation run (total $510$ runs). %Purple circles represent scenarios where adversaries control at most $20\%$ of any layer. Green `+' signs correspond to simulations where  adversaries make up $40\%$ of at least one layer. %Figure (a) shows results for scenarios where all vanilla nodes act reliably, while Figure (b) shows results when a fraction of vanilla nodes act unreliably.
}
\label{fig:adv-rel}
\end{figure}

We run $510$ simulations with various combinations of $|A|$ adversaries and $|T|$ targets, set independently to values $1\leq |T|\leq 64$ and $1\leq |A|\leq 64$. 
In each experiment, we compute the adversarial cost for the $|A|$ malicious nodes $c_A = |A| - \sum_{i \in A} \hat{\rho}_i$ and the cost imposed on the $|T|$ targets $c_T = |T| - \sum_{j \in T} \hat{\rho}_j$. 
We show in Fig.~\ref{fig:adv-rel} a sample per experiment, situated according to the resulting target and adversarial costs $c_T$ and $c_A$. 
Note that the overall attack scale increases proportionally with $|A|\times|T|$, since this determines the number of links on which packets are dropped---and thus the potential magnitude of $c_A$ and $c_T$. 
%In our experiments the scale of $|A|$x$|T|$ ranges from $1$ (when there is a single adversary and a single target) to $1024$ (when there are $32$ adversaries and $32$ targets). 
This results in $c_A$ and $c_T$ values ranging from below $10^{-2}$ to above $10$. We use log scale to represent this wide range in Fig.~\ref{fig:adv-rel}.  
% scale attacks. 
%We use logarithmic axes to summarize in one plot the results of attacks at different scales. 
\begin{figure}[ht]
    \centering
\includegraphics[width=0.98\columnwidth]{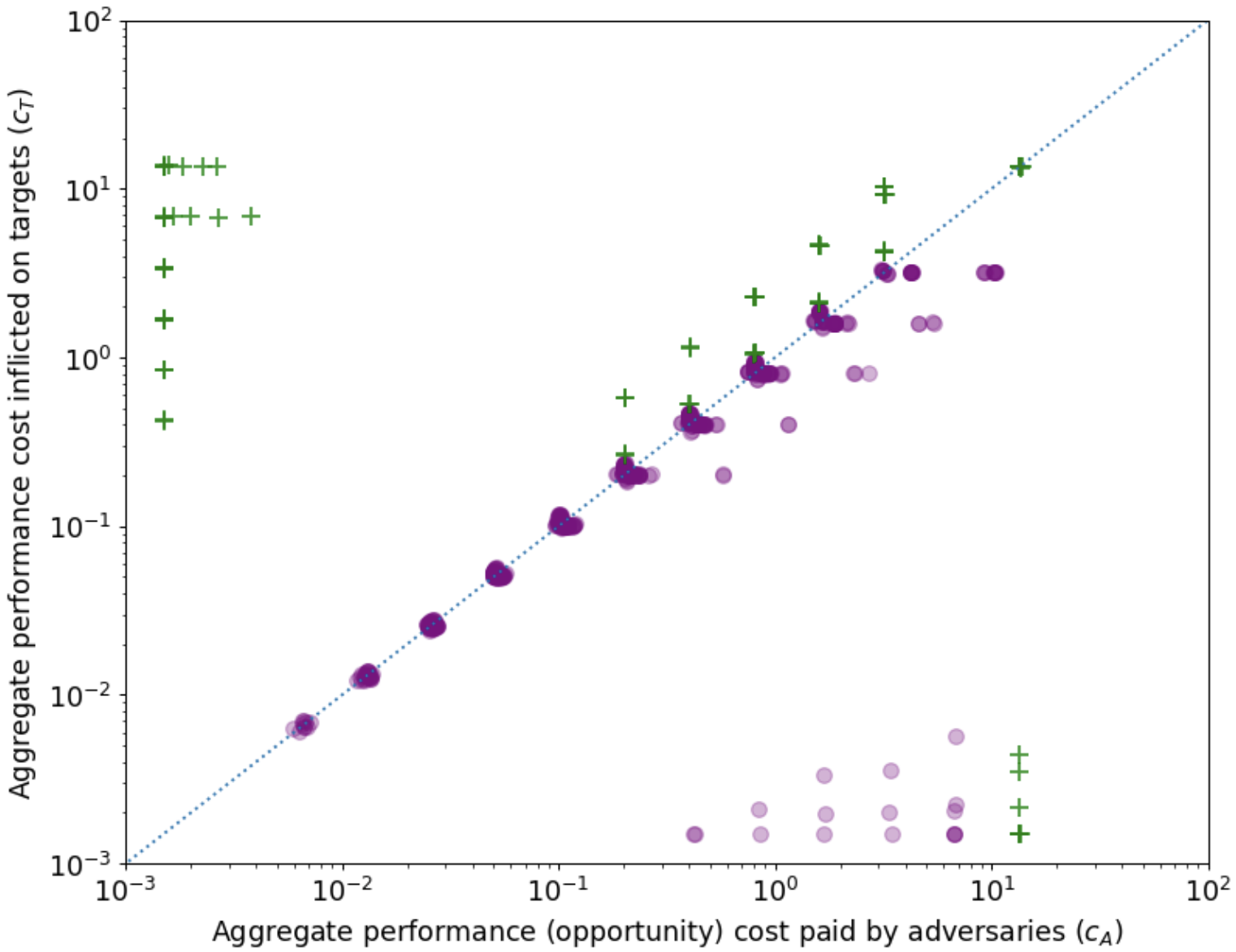}
    \caption{%Results for settings with both adversarial and unreliable nodes. In the scatter plot 
    Results for adversarial settings where some honest nodes are unreliable.
    }
    \label{fig:adv-unrel}
\end{figure}

Each simulation result is plotted as a point with coordinates $(c_A, c_T)$. 
%first for scenarios where all nodes have perfect reliability and all the packet drops are due to the attack (\ref{fig:adv-rel}), and second for scenarios where some nodes are unreliable and thus packet drops are due to a mix of random unreliable behavior and actual attack (\ref{fig:adv-unrel}). 
%There are two types of symbol in the figures: purple circles depict results of scenarios where adversaries control at most $20\%$ ($16$ nodes) of any layer; while green `+' symbols correspond to  scenarios where adversaries control $40\%$ ($32$ nodes) of at least one layer. 
As shown, the results cluster closely around the diagonal  $c_A = c_T$, indicating that, by engaging in the attack, the malicious nodes incur a combined reliability penalty roughly equal to the degradation they inflict on the target set---even in scenarios where $40\%$ of a target’s predecessors and/or successors are adversarial.
%In the experiments presented here, we assume that all honest nodes have perfect reliability and that all observed packet drops are due to adversarial activity. %Appendix~\ref{app-results-adversarial-unrel} provides additional results for mixed settings involving both adversarial and honest-but-unreliable nodes.

\noindent
{\bf Combining Adversarial and Unreliable nodes.}
\label{app-results-adversarial-unrel}
We finally examine scenarios where some nodes not involved in the attack are unreliable. Similarly to the %evaluation in Section~\ref{sec:perf-adversarial}, 
previous case, we compute the adversarial cost for $|A|$ malicious nodes ($1 \leq |A| \leq 64$) as $c_A = |A| - \sum_{i \in A} \hat{\rho}_i$ and the cost imposed on the $|T|$ targets ($1 \leq |T| \leq 64$) as $c_T = |T| - \sum_{j \in T} \hat{\rho}_j$. 
We run $630$ simulations. Each simulation run produces a result sample $(c_A, c_T)$ that we represent as a dot in a scatter plot. We represent scenarios with large number of adversaries with a different symbol to better evaluate the impact of high corruption rates. 
Purple circles represent scenarios where adversaries control at most $20\%$ ($16$ out of $80$ nodes) of any layer. Green `+' signs correspond to simulations where  adversaries make up $40\%$ ($32$ out of $80$ nodes) of at least one layer. 
The results are shown in Fig.~\ref{fig:adv-unrel}, where we can see that adversary and target penalties are still symmetric in the majority of scenarios. We can see that $c_A \approx c_T$ in all scenarios below a certain scale.
%; e.g., when $A$x$T = 16$, $c_A, c_T \approx 0.1$ for all simulation runs. 
As the attack scale (proportional to $|A|$x$|T|$) increases however, there are increased chances that unreliable nodes affect the measured reliability of either adversaries or targets, making the attack cost asymmetric. This happens in particular when the combination of attack-related drops and unreliability affects the median incoming or outgoing link reliability of a node, i.e., when more than half of a node's incoming or outgoing links have sub-par performance for one or another reason.
%, as this causes the node to be labeled as \textit{unreliable}. 
This makes the node be labeled as \textit{unreliable} and increases the attribution of drops to that node. This can more easily happen to targets that are under attack from a large number of adversaries in an adjacent layer, as well as to adversaries that attack a large number of targets. 
%On the left side of Figure~\ref{fig:adv-unrel} above the diagonal we see a set of results for which the adversary succeeds in causing a large reliability degradation on targets at no penalty to itself ($c_A \ll c_T$). This degree of adversarial success requires controlling $40\%$ ($32$ out of $80$ nodes) of at least one layer (all those samples are green `+' symbols) but also some luck for the adversary. In various other instances with the same degree of adversarial control we observe that either: (i) the adversary has a penalty $c_A$ that is lower than that of the targets' $c_T$, but still proportional to it and non-negligible (row of green `+' symbols over the diagonal for which $c_A < c_T$); (ii) the adversary's penalty is symmetric with the target's (`+' symbols on the diagonal for which $c_A \approx c_T$); or (iii) in a few cases the adversary pays a large penalty $c_A > 10$ without inflicting any penalty on the targets (`+' symbols on the bottom of the plot for which $c_A \gg c_T$). 
The considered level of unreliability is not helpful for adversaries that control up to $20\%$ of any layer (purple circle symbols). In these cases, at best the adversary pays a symmetric penalty (results on the diagonal, for which $c_A \approx c_T$), while in other instances it pays a higher cost than the target (circles below the diagonal for which $c_A > c_T$). In some instances the adversary pays a large penalty without the attack having an effect on the target (samples on the bottom of the figure for which $c_A \gg c_T$).

\section{Related Work}
\label{sec:related}
%mixnets

%First introduced by Chaum~\cite{chaum1981untraceable} in the early 1980s, mixnets are a key technique for anonymously routing packets that has inspired multiple lines of research that target different use cases for mixnets, each with its own tradeoffs. 

%such that, as long as packets pass through at least one honest node, they cannot be traced from source to destination. This is the case even if the adversary can eavesdrop on all network communications, as each mix node's input and output packets are unlinkable both in terms of content and timing. 

\noindent
\textbf{Mixnets for voting.}
A key use case for mixnets in the literature is electronic voting~\cite{DBLP:conf/ccs/Neff01, DBLP:conf/asiacrypt/Wikstrom05, haines2020sok}, where the goals are to ensure both unlinkability between voters and their ballots, and the verifiability of the final tally. 
In such applications, mixnets must offer strong integrity guarantees: each client is authorized to send exactly one well-formed ballot (typically authorized by some form of credential), and mix nodes are required to prove that they correctly process every packet without dropping, injecting, altering, or substituting any messages.  
These guarantees are achieved by routing all transmissions through a public broadcast channel and requiring per-batch NIZK shuffle proofs~\cite{krips2021more} to demonstrate that mix node outputs are valid permutations of their inputs. This supports high-assurance verifiability but introduces latency and communication costs only acceptable in voting applications.

%, where such per-packet overheads are justified. 
% the batch-size is large and votes can be tallied within . % which is a notably latency-tolerant use case where, due to the stakes involved, a significant overhead per packet (ballot) is justified. Such communication latencies and broadcast overheads are however impractical for any other applications. 

\noindent
\textbf{Mixnets for messaging.} 
%Micro-blogging is a latency-tolerant application that only requires occasional broadcast of a short message. 
%Atom~\cite{kwon2017atom} proposes a mixnet for this use case that %uses interconnected groups of mix nodes operating in parallel to achieve scalability. It proposes a construction that 
%relies on lightweight traps rather than NIZKs to increase efficiency  while protecting against active attacks by malicious mix nodes. 
Early mixnets were designed for high-latency messaging (email) and implicitly assumed that mix nodes were reliable~\cite{chaum1981untraceable, mixmaster, danezis2003mixminion}. %Mixminion~\cite{danezis2003mixminion} relying on centralized directory authorities but with minimal co-ordination between mix nodes. 
%More recently, cMix~\cite{chaum2017cmix} introduced mechanisms to reduce message transmission latency by moving time-consuming public key operations to a pre-computation phase. As a result, cMix is able to route messages with sub-second latency. However, this reliance on pre-computation and cascades makes cMix particularly vulnerable to server failures: if even a single node in the cascade fails during either the pre-computation or transmission phases, clients sending a batch of messages must restart the process entirely.
Most modern designs, including  cMix~\cite{chaum2017cmix}, Vuvuzela ~\cite{van2015vuvuzela}, Stadium~\cite{tyagi2017stadium}, and Groove~\cite{groove} similarly assume highly reliable servers and do not incorporate mechanisms to estimate node reliability.
%Stadium~\cite{tyagi2017stadium} relies on verifiable shuffles for parallel mixnets that can scale to millions of users with latency still within minutes. Groove~\cite{groove} introduces oblivious delegation mechanisms to allow clients to go offline without compromising their privacy properties and latency of half a minute for a million users. 
% The earliest design proposing a low-latency general purpose mixnet is  ISDN-MIXes~\cite{ISDN-mixes}, which assumes a batch-based mix cascade topology and thus becomes unavailable if one of the mix servers goes offline. 
%Vuvuzela, built for private messaging, is able to route 240-byte messages with high throughput and a latency of more than half a minute. It assumes that all servers are perfectly reliable: since it relies on a chain of servers, any one server being offline would render the messaging service unavailable to all users.
A few proposals do consider reliability properties. The Atom mixnet~\cite{kwon2017atom} uses lightweight ``trap'' ciphertexts to detect faulty behavior, but its latency (28 minutes for one million 32-byte messages) scales with message number and size. XRD~\cite{XRD} improves on Atom's scalability with aggregate hybrid shuffles and identifies misbehaving servers via NIZKs, but its latency scales linearly with the number of Byzantine users. Karaoke~\cite{karaoke}, the fastest in this category (6.8s for two million clients), uses Bloom filters to detect dropped messages but treats all drops as adversarial and otherwise assumes perfectly reliable servers; its focus is on deanonymization effects of unreliability rather than on scoring node reliability.
Across these systems~\cite{XRD, kwon2017atom, karaoke}, latency and broadcast overhead grow with client message volume. Our scheme instead runs reliability estimation \textit{a posteriori}, leaving packet-forwarding latency unaffected, and keeps broadcast overhead constant for a fixed topology and target accuracy, regardless of traffic volume.

%For example, although Atom's amount of broadcast data is not explicitly provided~\cite{kwon2017atom}, given its design, 
 %The Vuvuzela design does not discuss any mechanisms to measure whether any messages are dropped or substituted by intermediate servers, or provide any solutions to cope with a server going down. 
%% Stadium
%, but the latency of messages is still in the order of minutes, and measuring reliability is out of the scope of Stadium. 

%% Groove
%  Groove's message latency is half a minute for one million users, and it also increases with the number of users. 
 
%in the order of minutes and the message latency grows with the number of client messages being routed per round.

% loopix 
\noindent
\textbf{Continuous-time mixnets.} 
Loopix~\cite{piotrowska2017loopix} is not \textit{round-based}: it processes packets individually with random per-hop delays~\cite{kesdogan1998stop,danezis-TA-continuous}, enabling clients to configure latency profiles~\cite{beta-mixing} for different applications. This makes Loopix well-suited for low-latency use, but it provides no node-reliability mechanism. The Nym mixnet~\cite{diaz2021nym}, based on Loopix, currently relies on centralized network monitors that loop packets through the mixnet to compute node scores; forthcoming work~\cite{cao-green-attack} highlights weaknesses of this centralized approach.

\noindent
\textbf{Mixnet reliability and accountability.} 
%Finally, we review prior work that specifically tackles reliability or accountability. 
%Early work on reliability includes the proposal by Danezis and Sassaman~\cite{danezis-heartbeat} of having mix nodes send \textit{heartbeat} traffic to themselves through the mixnet, similar to the \textit{loops} of cover traffic implemented in Loopix~\cite{piotrowska2017loopix}, to detect active attacks by malicious nodes that drop or substitute messages.  
% 
The idea of estimating a mix node’s reliability using a reputation score was first introduced by Dingledine \textit{et al.}~\cite{dingledine2001reputation}, who proposed a scheme relying on semi-trusted witnesses to verify whether nodes behave correctly. However, their model does not account for active adversaries attempting to manipulate reliability measurements. 
Subsequent work~\cite{dingledine2003reliable} eliminates the need for witnesses by using test measurements and failure reports, which are tied to reputation, to construct mix cascades that can detect packet drops or substitutions. 
This design, however, is vulnerable to so-called ``creeping death'' attacks, where adversaries strategically bias measurements to gradually erode the reputation of honest nodes more than that of malicious ones.
Our protocol addresses this limitation by ensuring that adversarial nodes incur reputational penalties that are proportional to the harm they inflict on their targets. 

More recent proposals~\cite{miranda,DBLP:conf/esorics/BoyenHM20} either treat any packet loss as malicious or rely on trusted auditors. The ``trip-wire'' technique from verifiable mixnets~\cite{haines2020sok}, used by~\cite{DBLP:conf/asiacrypt/KhazaeiMW12,DBLP:conf/esorics/BoyenHM20}, resembles our measurement packets in employing indistinguishable test packets, but yields only a binary outcome rather than a representative sample. Our VRF-based measurement packets, by contrast, enable quantitative link and node reliability estimation that reflects user experience.

\section{Conclusion}
\label{sec:conclusion}
% summary of achievements
We have presented the first decentralized, scalable, low-latency scheme for estimating the reliability of a mixnet's links and nodes. Our \textit{link reliability estimation} protocol relies on covert measurement packets and a novel \textit{VRF-based routing} primitive that enforces adherence to the mixnet's routing policy, preventing malicious clients from biasing paths. The resulting link estimates are combined into \textit{node reliability scores} that identify underperforming mix nodes with optimal communication overhead.

We empirically evaluated our scheme via simulation. In a Nym-like network with three routing layers and $80$ nodes per layer, 1 million measurement packets suffice to obtain accurate estimates in both unreliable and adversarial settings. 
A central insight is that, for a fixed topology and target accuracy, the required number of measurements remains constant as client traffic grows, making the protocol overhead independent of traffic volume.
Unlike heavy-weight verifiable mixnets, our design imposes no latency on packet delivery (verification runs \textit{a posteriori}) and only a small fraction of packets, with minimal per-packet information, ever reaches the broadcast channel. 
% AK: no need to get back to the below.. 
%---in contrast to full-packet replication at every hop.

% generalization to other mixnets and designs
Although designed for continuous-time mixnets, our protocols can also apply to batch-based systems, since they operate over epochs, and to alternative graph topologies as long as the routing policy is public. The number of links determines the number of measurements required: free routes are the most expensive, parallel cascades the cheapest. Free routes also yield the strongest node scoring, tolerating nearly half of all nodes being adversarial, because every node is predecessor and successor of every other node; cascades on the other hand cannot attribute link drops at all, since each node has only one predecessor and successor.

%%% AK: I think the below paragraph can be removed ...  it is too detailed/specific for the conclusion. 
%Both the client and the gateway can generate measurements; what matters is that the process cannot be biased. VRF-based selection tied to packet counters enforces this --- measurements form a random sample of legitimate traffic, and public verification of no-skipping is possible. If generation were independent of credential-bound counters, malicious clients or gateways could over-generate or omit measurements to frame nodes and bias estimates. Since gateways are typically online while clients may have intermittent connectivity, having gateways submit measurements is the more practical choice.

In terms of future work, interesting  directions include further exploring the trade-off between the utility-maximizing and fully-malicious settings with respect to communication overhead, and analyzing \textit{griefing attacks} on node reliability estimation, where an adversary sacrifices its own reliability to harm honest nodes.

\subsection*{Acknowledgments.}

The authors would like to thank Leif Ryge for early ideas of `secret shopper' traffic to measure the  performance for Tor, and further discussions with Jeff Burdges and Leif Ryge for applying this idea in the context of mixnets. We'd like to thank Dan Boneh and Nym advisors Bart Preneel, George Danezis, and Ben Laurie for discussions, as well anonymous reviewers for extensive comments. This research was supported in part by CyberSecurity Research Flanders with reference number VR20192203 and by Nym Technologies SA. 

\noindent
\textbf{Use of AI-based tools:} 
ChatGPT and Claude were used in the final stages of this paper's editing, to improve the readability and flow of the text, and to correct typos. 

\bibliographystyle{IEEEtran}
%\bibliographystyle{ACM-Reference-Format}
%\balance
\bibliography{mybib}

@article{cao-green-attack,
	author = {Cao, Xinmu Alexis and Green, Matthew},
	journal = {PoPETs (to appear)},
	title = {Analysis and Attacks on the Reputation System of Nym},
	volume = {2026},
	year = {2026},
    issue = {2}, 
    }

@inproceedings{merkle1988,
  author    = {Ralph C. Merkle},
  title     = {A Digital Signature Based on a Conventional Encryption Function},
  booktitle = {Advances in Cryptology -- CRYPTO '87},
  series    = {Lecture Notes in Computer Science},
  volume    = {293},
  pages     = {369--378},
  publisher = {Springer},
  year      = {1988}
}

@inproceedings{DBLP:conf/asiacrypt/KhazaeiMW12,
	author = {Shahram Khazaei and Tal Moran and Douglas Wikstr{\"{o}}m},
	bibsource = {dblp computer science bibliography, https://dblp.org},
	booktitle = {Advances in Cryptology - {ASIACRYPT} 2012 - 18th International Conference on the Theory and Application of Cryptology and Information Security, Beijing, China, December 2-6, 2012. Proceedings},
	doi = {10.1007/978-3-642-34961-4\_37},
	editor = {Xiaoyun Wang and Kazue Sako},
	pages = {607--625},
	publisher = {Springer},
	series = {Lecture Notes in Computer Science},
	title = {A Mix-Net from Any {CCA2} Secure Cryptosystem},
	url = {https://doi.org/10.1007/978-3-642-34961-4\_37},
	volume = {7658},
	year = {2012}}

@inproceedings{haines2020sok,
	author = {Haines, Thomas and M{\"u}ller, Johannes},
	booktitle = {2020 IEEE 33rd Computer Security Foundations Symposium (CSF)},
	organization = {IEEE},
	pages = {49--64},
	title = {SoK: techniques for verifiable mix nets},
	year = {2020}}

@article{davidson2018privacy,
	author = {Davidson, Alex and Goldberg, Ian and Sullivan, Nick and Tankersley, George and Valsorda, Filippo},
	journal = {Proceedings on Privacy Enhancing Technologies},
	title = {Privacy pass: Bypassing internet challenges anonymously},
	year = {2018}}

@inproceedings{greubel2020,
	address = {New York, NY, USA},
	author = {Greubel, Andre and Pohl, Steffen and Kounev, Samuel},
	booktitle = {Proceedings of the 36th Annual Computer Security Applications Conference},
	doi = {10.1145/3427228.3427238},
	isbn = {9781450388580},
	location = {Austin, USA},
	numpages = {12},
	pages = {129--140},
	publisher = {Association for Computing Machinery},
	series = {ACSAC '20},
	title = {Quantifying measurement quality and load distribution in Tor},
	url = {https://doi.org/10.1145/3427228.3427238},
	year = {2020}}

@inproceedings{zhang2021tnras,
	author = {Zhang, Wenzhen and Lu, Tianbo and Du, Zeyu},
	booktitle = {Proceedings of the 2021 11th International Conference on Communication and Network Security},
	pages = {21--26},
	title = {TNRAS: Tor nodes reliability analysis scheme},
	year = {2021}}

@inproceedings{kate2023flexirand,
	author = {Kate, Aniket and Mangipudi, Easwar Vivek and Maradana, Siva and Mukherjee, Pratyay},
	booktitle = {Proceedings of the 2023 ACM SIGSAC Conference on Computer and Communications Security},
	pages = {1776--1790},
	title = {Flexirand: Output private (distributed) vrfs and application to blockchains},
	year = {2023}}

@article{beta-mixing,
	author = {Guirat, Iness and Das, Debajyoti and Diaz, Claudia},
	doi = {10.56553/popets-2024-0059},
	journal = {PoPETs},
	month = {04},
	pages = {464-478},
	title = {Blending Different Latency Traffic With Beta Mixing},
	volume = {2024},
	year = {2024}}

@inproceedings{danezis-TA-continuous,
	author = {Danezis, George},
	booktitle = {Privacy Enhancing Technologies (PETS)},
	doi = {10.1007/11423409_3},
	isbn = {3540262032},
	numpages = {16},
	pages = {35--50},
	publisher = {Springer},
	title = {The traffic analysis of continuous-time mixes},
	url = {https://doi.org/10.1007/11423409_3},
	year = {2004}}

@inproceedings{karaoke,
	author = {David Lazar and Yossi Gilad and Nickolai Zeldovich},
	booktitle = {Symposium on Operating Systems Design and Implementation (OSDI)},
	isbn = {978-1-939133-08-3},
	pages = {711--725},
	publisher = {USENIX Association},
	title = {Karaoke: Distributed Private Messaging Immune to Passive Traffic Analysis},
	url = {https://www.usenix.org/conference/osdi18/presentation/lazar},
	year = {2018}}

@inproceedings{groove,
	author = {Ludovic Barman and Moshe Kol and David Lazar and Yossi Gilad and Nickolai Zeldovich},
	booktitle = {16th USENIX Symposium on Operating Systems Design and Implementation (OSDI 22)},
	isbn = {978-1-939133-28-1},
	pages = {735--750},
	publisher = {USENIX Association},
	title = {Groove: Flexible {Metadata-Private} Messaging},
	url = {https://www.usenix.org/conference/osdi22/presentation/barman},
	year = {2022}}

@inproceedings{XRD,
	author = {Kwon, Albert and Lu, David and Devadas, Srinivas},
	booktitle = {Usenix Conference on Networked Systems Design and Implementation},
	isbn = {9781939133137},
	pages = {759-776},
	publisher = {USENIX Association},
	series = {NSDI},
	title = {{XRD}: scalable messaging system with cryptographic privacy},
	year = {2020}}

@inproceedings{kesdogan1998stop,
	author = {Kesdogan, Dogan and Egner, Jan and B{\"u}schkes, Roland},
	booktitle = {Information Hiding},
	organization = {Springer},
	pages = {83--98},
	title = {Stop-and-go-mixes providing probabilistic anonymity in an open system},
	year = {1998}}

@inproceedings{miranda,
	author = {Hemi Leibowitz and Ania M. Piotrowska and George Danezis and Amir Herzberg},
	booktitle = {USENIX Security Symposium},
	isbn = {978-1-939133-06-9},
	pages = {1841--1858},
	title = {No Right to Remain Silent: Isolating Malicious Mixes},
	url = {https://www.usenix.org/conference/usenixsecurity19/presentation/leibowitz},
	year = {2019}}

@article{mixmaster,
	author = {Cottrell, Lance and Palfrader, Peter and Sassaman, Len},
	journal = {Online specification},
	title = {Mixmaster Protocol Version 2< draft-moeller-v2-01.txt},
	year = {2003}}

@article{Diaz2022Reward-mixnet,
	author = {Diaz, Claudia and Halpin, Harry and Kiayias, Aggelos},
	journal = {Cryptoeconomic Systems},
	month = {June},
	number = {1},
	title = {Reward {Sharing} for {Mixnets}},
	volume = {2},
	year = {2022}}

@inproceedings{sphinx-2009,
	author = {George Danezis and Ian Goldberg},
	bibsource = {dblp computer science bibliography, https://dblp.org},
	booktitle = {{IEEE} Symposium on Security and Privacy (S{\&}P 2009)},
	doi = {10.1109/SP.2009.15},
	pages = {269--282},
	publisher = {{IEEE} Computer Society},
	title = {Sphinx: {A} Compact and Provably Secure Mix Format},
	url = {https://doi.org/10.1109/SP.2009.15},
	year = {2009}}

@inproceedings{DBLP:conf/asiacrypt/Wikstrom05,
	author = {Douglas Wikstr{\"{o}}m},
	bibsource = {dblp computer science bibliography, https://dblp.org},
	booktitle = {Advances in Cryptology - {ASIACRYPT}},
	doi = {10.1007/11593447\_15},
	pages = {273--292},
	publisher = {Springer},
	series = {LNCS},
	title = {A Sender Verifiable Mix-Net and a New Proof of a Shuffle},
	url = {https://doi.org/10.1007/11593447\_15},
	volume = {3788},
	year = {2005}}

@inproceedings{DBLP:conf/ccs/Neff01,
	author = {C. Andrew Neff},
	bibsource = {dblp computer science bibliography, https://dblp.org},
	booktitle = {Conference on Computer and Communications Security, {CCS}},
	doi = {10.1145/501983.502000},
	pages = {116--125},
	publisher = {{ACM}},
	title = {A verifiable secret shuffle and its application to e-voting},
	url = {https://doi.org/10.1145/501983.502000},
	year = {2001}}

@inproceedings{tyagi2017stadium,
	author = {Tyagi, Nirvan and Gilad, Yossi and Leung, Derek and Zaharia, Matei and Zeldovich, Nickolai},
	booktitle = {Operating Systems Principles},
	pages = {423--440},
	title = {Stadium: A distributed metadata-private messaging system},
	year = {2017}}

@inproceedings{van2015vuvuzela,
	author = {Van Den Hooff, Jelle and Lazar, David and Zaharia, Matei and Zeldovich, Nickolai},
	booktitle = {Operating Systems Principles},
	pages = {137--152},
	title = {Vuvuzela: Scalable private messaging resistant to traffic analysis},
	year = {2015}}

@article{chaum1981untraceable,
	author = {Chaum, David L},
	journal = {Communications of the ACM},
	number = {2},
	pages = {84--90},
	publisher = {ACM New York, NY, USA},
	title = {Untraceable electronic mail, return addresses, and digital pseudonyms},
	volume = {24},
	year = {1981}}

@misc{diaz2021nym,
	author = {Diaz, Claudia and Halpin, Harry and Kiayias, Aggelos},
	howpublished = {Whitepaper of Nym Technologies SA, version 1.0},
	month = {February},
	pages = {38},
	title = {The {N}ym {N}etwork},
	year = {2021}}

@inproceedings{borisov2007denial,
	author = {Borisov, Nikita and Danezis, George and Mittal, Prateek and Tabriz, Parisa},
	booktitle = {Proceedings of the 14th ACM conference on Computer and communications security},
	pages = {92--102},
	title = {Denial of service or denial of security?},
	year = {2007}}

@inproceedings{krips2021more,
	author = {Krips, Toomas and Lipmaa, Helger},
	booktitle = {CT-RSA},
	organization = {Springer},
	pages = {252--275},
	title = {More efficient shuffle argument from unique factorization},
	year = {2021}}

@inproceedings{piotrowska2017loopix,
	author = {Piotrowska, Ania M and Hayes, Jamie and Elahi, Tariq and Meiser, Sebastian and Danezis, George},
	booktitle = {USENIX Security Symposium},
	pages = {1199--1216},
	title = {The {L}oopix anonymity system},
	year = {2017}}

@inproceedings{kwon2017atom,
	author = {Kwon, Albert and Corrigan-Gibbs, Henry and Devadas, Srinivas and Ford, Bryan},
	booktitle = {Symposium on Operating Systems Principles},
	organization = {ACM},
	pages = {406--422},
	title = {Atom: Horizontally scaling strong anonymity},
	year = {2017}}

@inproceedings{danezis2003mixminion,
	author = {Danezis, George and Dingledine, Roger and Mathewson, Nick},
	booktitle = {IEEE Symposium on Security and Privacy},
	pages = {2--15},
	title = {Mixminion: Design of a type {III} anonymous remailer protocol},
	year = {2003}}

@inproceedings{dingledine2004tor,
	author = {Dingledine, Roger and Mathewson, Nick and Syverson, Paul},
	booktitle = {{USENIX} Security Symposium},
	pages = {303--320},
	title = {Tor: The second-generation onion router},
	year = {2004}}

@inproceedings{DBLP:conf/focs/MicaliRV99,
	author = {Silvio Micali and Michael O. Rabin and Salil P. Vadhan},
	bibsource = {dblp computer science bibliography, https://dblp.org},
	booktitle = {Foundations of Computer Science, {FOCS}},
	doi = {10.1109/SFFCS.1999.814584},
	pages = {120--130},
	publisher = {{IEEE} Computer Society},
	title = {Verifiable Random Functions},
	url = {https://doi.org/10.1109/SFFCS.1999.814584},
	year = {1999}}

@inproceedings{dingledine2001reputation,
	author = {Dingledine, Roger and Freedman, Michael J and Hopwood, David and Molnar, David},
	booktitle = {Information Hiding},
	organization = {Springer},
	pages = {126--141},
	title = {A reputation system to increase MIX-net reliability},
	year = {2001}}

@inproceedings{chaum2017cmix,
	author = {Chaum, David and Das, Debajyoti and Javani, Farid and Kate, Aniket and Krasnova, Anna and De Ruiter, Joeri and Sherman, Alan T},
	booktitle = {{ACNS}},
	organization = {Springer},
	pages = {557--578},
	title = {cMix: Mixing with minimal real-time asymmetric cryptographic operations},
	year = {2017}}

@inproceedings{dingledine2003reliable,
	author = {Dingledine, Roger and Syverson, Paul},
	booktitle = {Financial Cryptography (FC)},
	organization = {Springer},
	pages = {253--268},
	title = {Reliable MIX cascade networks through reputation},
	year = {2003}}

@article{DBLP:journals/popets/RialP23,
	author = {Alfredo Rial and Ania M. Piotrowska},
	bibsource = {dblp computer science bibliography, https://dblp.org},
	doi = {10.56553/POPETS-2023-0116},
	journal = {PoPETs},
	number = {4},
	pages = {381--415},
	title = {Compact and Divisible E-Cash with Threshold Issuance},
	url = {https://doi.org/10.56553/popets-2023-0116},
	volume = {2023},
	year = {2023}}

@inproceedings{DBLP:conf/crypto/CamenischS03,
	author = {Jan Camenisch and Victor Shoup},
	bibsource = {dblp computer science bibliography, https://dblp.org},
	booktitle = {Advances in Cryptology - {CRYPTO} 2003, 23rd Annual International Cryptology Conference, Santa Barbara, California, USA, August 17-21, 2003, Proceedings},
	doi = {10.1007/978-3-540-45146-4\_8},
	editor = {Dan Boneh},
	pages = {126--144},
	publisher = {Springer},
	series = {Lecture Notes in Computer Science},
	title = {Practical Verifiable Encryption and Decryption of Discrete Logarithms},
	url = {https://doi.org/10.1007/978-3-540-45146-4\_8},
	volume = {2729},
	year = {2003}}

@inproceedings{DBLP:conf/sp/BunzBBPWM18,
	author = {Benedikt B{\"{u}}nz and Jonathan Bootle and Dan Boneh and Andrew Poelstra and Pieter Wuille and Gregory Maxwell},
	bibsource = {dblp computer science bibliography, https://dblp.org},
	booktitle = {2018 {IEEE} Symposium on Security and Privacy, {SP} 2018, Proceedings, 21-23 May 2018, San Francisco, California, {USA}},
	doi = {10.1109/SP.2018.00020},
	pages = {315--334},
	publisher = {{IEEE} Computer Society},
	title = {Bulletproofs: Short Proofs for Confidential Transactions and More},
	url = {https://doi.org/10.1109/SP.2018.00020},
	year = {2018}}

@misc{bulletproofsplusplus,
	author = {Liam Eagen and Sanket Kanjalkar and Tim Ruffing and Jonas Nick},
	howpublished = {Cryptology ePrint Archive, Paper 2022/510},
	note = {\url{https://eprint.iacr.org/2022/510}},
	title = {Bulletproofs++: Next Generation Confidential Transactions via Reciprocal Set Membership Arguments},
	url = {https://eprint.iacr.org/2022/510},
	year = {2022}}

@inproceedings{DBLP:conf/pkc/DodisY05,
	author = {Yevgeniy Dodis and Aleksandr Yampolskiy},
	bibsource = {dblp computer science bibliography, https://dblp.org},
	booktitle = {Public Key Cryptography ({PKC})},
	doi = {10.1007/978-3-540-30580-4\_28},
	pages = {416--431},
	publisher = {Springer},
	series = {LNCS},
	title = {A Verifiable Random Function with Short Proofs and Keys},
	url = {https://doi.org/10.1007/978-3-540-30580-4\_28},
	volume = {3386},
	year = {2005}}

@article{bloom70,
	address = {New York, NY, USA},
	author = {Bloom, Burton H.},
	doi = {10.1145/362686.362692},
	issn = {0001-0782},
	issue_date = {July 1970},
	journal = {Commun. ACM},
	month = {jul},
	number = {7},
	numpages = {5},
	pages = {422--426},
	publisher = {Association for Computing Machinery},
	title = {Space/time trade-offs in hash coding with allowable errors},
	url = {https://doi.org/10.1145/362686.362692},
	volume = {13},
	year = {1970}}

@inproceedings{DBLP:conf/fc/FurukawaS06,
	author = {Jun Furukawa and Kazue Sako},
	bibsource = {dblp computer science bibliography, https://dblp.org},
	booktitle = {Financial Cryptography and Data Security},
	doi = {10.1007/11889663\_8},
	pages = {111--125},
	publisher = {Springer},
	series = {LNCS},
	title = {An Efficient Publicly Verifiable Mix-Net for Long Inputs},
	url = {https://doi.org/10.1007/11889663\_8},
	volume = {4107},
	year = {2006}}

@inproceedings{DBLP:conf/crypto/Wagner02,
	author = {David A. Wagner},
	bibsource = {dblp computer science bibliography, https://dblp.org},
	booktitle = {Advances in Cryptology, {CRYPTO}},
	doi = {10.1007/3-540-45708-9\_19},
	pages = {288--303},
	publisher = {Springer},
	series = {LNCS},
	title = {A Generalized Birthday Problem},
	url = {https://doi.org/10.1007/3-540-45708-9\_19},
	volume = {2442},
	year = {2002}}

@inproceedings{DBLP:conf/esorics/BoyenHM20,
	author = {Xavier Boyen and Thomas Haines and Johannes M{\"{u}}ller},
	bibsource = {dblp computer science bibliography, https://dblp.org},
	booktitle = {{ESORICS}},
	doi = {10.1007/978-3-030-59013-0\_17},
	pages = {336--356},
	publisher = {Springer},
	series = {LNCS},
	title = {A Verifiable and Practical Lattice-Based Decryption Mix Net with External Auditing},
	url = {https://doi.org/10.1007/978-3-030-59013-0\_17},
	volume = {12309},
	year = {2020}}

@article{clopper-pearson,
	author = {Clopper, C. J. and Pearson, E. S.},
	doi = {10.1093/biomet/26.4.404},
	eprint = {https://academic.oup.com/biomet/article-pdf/26/4/404/823407/26-4-404.pdf},
	issn = {0006-3444},
	journal = {Biometrika},
	month = {12},
	number = {4},
	pages = {404-413},
	title = {The Use of Confidence OR Fiducial Limits Illustrated in the case of the Binomal},
	url = {https://doi.org/10.1093/biomet/26.4.404},
	volume = {26},
	year = {1934}}

@inproceedings{DBLP:conf/pkc/CatalanoF13,
  author       = {Dario Catalano and
                  Dario Fiore},
  editor       = {Kaoru Kurosawa and
                  Goichiro Hanaoka},
  title        = {Vector Commitments and Their Applications},
  booktitle    = {Public-Key Cryptography - {PKC} 2013 - 16th International Conference
                  on Practice and Theory in Public-Key Cryptography, Nara, Japan, February
                  26 - March 1, 2013. Proceedings},
  series       = {Lecture Notes in Computer Science},
  pages        = {55--72},
  publisher    = {Springer},
  year         = {2013},
  url          = {https://doi.org/10.1007/978-3-642-36362-7\_5},
  doi          = {10.1007/978-3-642-36362-7\_5},
  bibsource    = {dblp computer science bibliography, https://dblp.org}
}

%\appendix

\appendices

\section{Proof of Theorem~\ref{thm:vrf}}
\label{app:security-theorem}

For reference we recall our notations in Table~\ref{tab:notation} and our underlying packet construction as illustrated in Figure~\ref{fig:dependencies-2gen}. 

\begin{proof} 
(I) 
%%%%%%%%%%%%
%
The key part in  proving the statement is  to demonstrate that 
the values used for the underlying mixnet encoding $\tilde r_i$ are indistinguishable 
to random values that the client would select independently.
The mechanism we use for the special headers 
is an extension of the Sphinx encoding and hence its security
can be argued as in \cite{sphinx-2009}. For completeness we provide the main argument below. 
The main dependency is
on the Decisional Diffie Hellman assumption and the random oracle model. 
For simplicity we argue first only the case that a single credential is used to 
send a single packet via a gateway. Consider the $i$-th hop of that packet. 
This involves the values $\alpha_i = \alpha_{i-1}^{b_{i-1}}$
and $y_i$, the public-key of the $i$-th node.  

We show how to incorporate
a challenge for the Decisional Diffie Hellman assumption $\langle g, a, b, c\rangle$ into the encoding calculation. We set the credential key $\alpha$ to be  $a$ and the node public-key $y_i$ to be equal to $b$. 
It follows that the value $\tilde r_i$
can be calculated as a function of the DDH challenge by $H(\mathsf{rnd}, c^e)$
where $e = b_0\ldots b_{i-1} \cdot r_\mathrm{pkt}$. 
Conditional on $e\neq 0$, it follows that $\tilde r_i$ is uniformly random, 
unless the adversary queries the value $c^e$ to the random oracle. 
Note that this event can only happen with non-negligible probability
in the case the challenge $\langle g, a, b, c\rangle$  is a DDH tuple 
(as in the other case, the value $c$ is random and hence can only be predicted with negligible probability by the adversary). 
As a result, if the adversary queries $c^e$ we can readily build a distinguisher by  examining the random oracle queries of the adversary. Specifically, the distinguisher can test each query $q$ for $q = c^{e}$ and output $1$ if this is the case while producing $0$ otherwise. It is easy to see that this algorithm will produce $1$ with non-negligible probability when the tuple $\langle g, a, b, c\rangle$ follows the DDH distribution while it will produce $0$ with overwhelming probability whenever 
$\langle g, a, b, c\rangle$  is a random tuple. Finally, observe that the probability that $e=0$ is negligible given the randomness of the constituent values in the random oracle model. 

It follows that the special header value $\alpha$ can be simulated independent of the packet and its payload under the pseudorandomness of the VRF, the DDH and in the random oracle model, hence as a result the only advantage gained by the adversary can be drawn by the underlying encoding which is assumed to be secure. 
Finally, note that the above argument generalizes via a hybrid argument in a straightforward manner to the case that there are multiple clients sending multiple packets. 
%% NO VRF                                         
%%Note that it is in this step that the VRF property is needed since we want to ensure that the $\alpha_0$ values of different packets appear independently sampled from the base group. The pseudorandomness property %of the underlying function  provides us exactly that: the value $r_\mathsf{pkt}$ would randomize $\alpha$ within the base group enabling us to repeat the DDH distinguishing argument for each packet. 
%

(II) The argument is similar to case (I). Recall our encoding mechanism introduces the values $\alpha_0, \alpha_1, 
\ldots, \alpha_\nu$ that accompany each packet when they are sent over the network. As mentioned already, this is a sequence of blinded values that follow the  Sphinx construction and hence the privacy proof would be identical. For completeness, the key argument is the following. Consider the two values $\alpha^*_0, \alpha^*_1$ originating from the packet processing performed by the node. We consider the two cases in the theorem statement. When the packets are processed it holds that 
there are two exponents $e_0,e_1$ such that  $(\alpha^*_0, \alpha^*_1) = (\alpha_0^{e_0}, \alpha_1^{e_1})$. The indistinguishability of $(\alpha^*_b, \alpha^*_{1-b})$ 
to $(\alpha^*_{1-b}, \alpha^*_{b})$
boils down to the pseudorandomness of the $e_0,e_1$ exponents which 
can easily  argued in a similar way as above under   DDH and the random oracle model.

(III) The statement asks for the pseudorandomness 
of the values $r_0,\ldots,r_{\nu-1}$, even in the case
that the values $vk, \alpha, y_0$ are adversarially chosen. 
Note that in this case we cannot rely on the DDH assumption or the properties of the VRF. Instead the key observation is that each value $r_i$ depends on the public nonce value via the random oracle as well as some other value $s_i$ that as long as it is unique, $r_i$ would be an independently sampled value. To prove this we consider the event that there are two values $s = g^{e}, s'=g^{e'}$  in this sequence  that are equal, i.e., $e=e'$. Depending on the step when these values have been calculated the exponents will be a product of a number of values some of which are committed by the adversary, e.g., the discrete logarithm of the key of an adversarial gateway, the discrete logarithm of a credential's randomization key $\alpha$ as well as values produced by random oracle evaluations of such values e.g., the blinding exponents $b_0, b_1, b_2,\ldots$ and the keys of the mix-nodes. Conditional on no collisions in the random oracle $H(\cdot)$ or the VRF function, all these individual values involved in $e,e'$ are distinct and, in case they are not outputs of the random oracle, are committed prior to the production of $\mathsf{nonce}$. Based on this, it follows that the event $e=e'$ would imply the event that 
$z_1 \cdot z_k \cdot h_1 \ldots h_t = 1$ where $h_1,\ldots,h_t$ are random oracle outputs that depend on $\mathsf{nonce}$ and $z_1,\ldots, z_k$ values selected by the adversary. It is easy to see via generalized birthday problem \cite{DBLP:conf/crypto/Wagner02} that this would require subexponential overhead in the worst-case (note in our setting $t$ is a small constant). 

(IV) This statement follows easily due to the fact that the only difference between measurement and non-measurement packets is in the calculation of the $\alpha_0$ value: in the case of a non-measurement packet this value is equal to $\alpha^{r_\mathsf{pkt}}$, while in the case of a measurement packet it is equal to $g^{r_\mathsf{pkt}}$. In the view of a relay node, the exponentiation by the (hidden) value  $r_\mathsf{pkt}$ results in a full randomization of the resulting element within the base group (with the precondition $\alpha\neq 1$); as a result the two cases are indistinguishable assuming the security of the VRF function. 
\end{proof}

\section{Approaches to setting $\hat{\beta}_e$}
\label{appendix-beta}

Here we propose and review three heuristic approaches to setting values for $\hat{\beta}_e$, which defines the fraction of drops ${d}^*_e$ in link $e=(i,j)$ that are attributed to successor $j$, while the remaining $(1-\hat{\beta}_e)\cdot {d}^*_e$ drops are attributed predecessor $i$. We assume reliable link transmissions such that %${\xi}_e = 0$, i.e., 
no packets are dropped in link $e=(i,j)$ if both $i$ and $j$ are online and functioning correctly. 

\subsection{Naive approach: $\hat{\beta}_e =1$}

A simple approach is to ``blame the receiver,'' which corresponds to setting $\hat{\beta}_e =1$ for all edges $e$. 
The rationale for this approach is that most of the hard work of processing a packet (deriving the decryption key and checking the packet tag for replay) is done by the time a node $i$ is able to store the packet tag in its tag-commitment. 
Thus, if a packet is lost in link $(i,j)$, chances are that the blame is with $j$, who failed to be online or otherwise to process the packet. 
This naive model performs quite well in scenarios where all link drops are due to random failures by \textit{honest but unreliable} nodes, since in practice chances are that the drop is due to the receiving end $j$ being either offline or congested. 

Such model is however easily subverted in a strategic adversarial setting: a malicious node $i$ can selectively drop received packets (after storing their tag in the tag-commitment) destined to one or more target successors $j$, in order to unfairly degrade their measured reliability $\hat{\rho}_j$. 
Given a mixnet with layer width $W$, an adversary that controls a set of $i \in A$ nodes in a layer can degrade the measured reliability $\hat{\rho}_j$ of any number of nodes $j$ in the succeeding layer, by up to $\frac{|A|}{W}$ at zero cost for the adversarial nodes in terms of lower measured reliability $\hat{\rho}_i$ for malicious nodes $i \in A$. 

%violating the measurement fairness property defined in Section~\ref{sec:problem}.
%\aggelos{mention here that this violates the fairness w.r.t. measurement property} 
 
\subsection{Symmetric approach: $\hat{\beta}_e =\frac{1}{2}$}

An alternative approach to penalize and discourage such adaptive malicious behaviour is to define node reliability considering that half the `blame' for drops in link $e=(i,j)$ is with $i$, and the other half with $j$, i.e., setting $\hat{\beta}_e =\frac{1}{2}$ for all links $e$. 
This symmetric attribution approach is effective at discouraging malicious packet drops as \textit{any} drop now affects the reliability of the adversarial node comparably to that of its target, which can be either a predecessor or a successor of the attacker. 
%We note that the symmetric attribution of drops however leads to an increasingly asymmetric reliability penalty for large scale attacks involving many adversarial nodes and many targets. In such cases, $\hat{\beta}_e=0.5$ $\forall e \in G$ begins to impose a larger reliability penalty on link successors than it does on predecessors. This is due to successors receiving significantly less total traffic (due to the large scale nature of the attack) and thus the attributed half of the drops represent a larger fraction of the total inputs than for the predecessors, to whom the other half of drops are attributed. 
%
A practical concern in cases where no attack takes place is that the symmetric attribution of blame results in a lowered reliability $\hat{\rho}_i$ for the predecessors $i$ of nodes $j$ that are either offline or congested and unable to process all the received traffic -- a scenario that may not be rare. This may unfairly degrade $\hat{\rho}_i < {\rho}_i$ for reliable nodes $i$ at the expense of overestimating $\hat{\rho}_j > {\rho}_j$ for unreliable nodes $j$.

Consider a case where $j \in F$ nodes in a layer of width $W$ go offline shortly after processing a few packets,\footnote{In the extreme case where a node $j$ has zero measurement packets in its tag-commitment for the epoch, i.e., $\sum_i {s}_{(i,j)}^* = 0$, then the node is marked as having reliability $\hat{\rho}_j = 0$, and all drops in the edges $(i,j)$ can be safely attributed to $j$, i.e., $\hat{\beta}_{(i,j)}=1$.} while all their predecessors $i$ are honest and reliable throughout the entire epoch, i.e., ${\rho}_i = 1$ for all nodes in the preceding layer. 
Attributing drops symmetrically causes $\hat{\rho}_i$ to be unfairly degraded by $\frac{|F|}{2W}$, while offline nodes $j$ unfairly get credit for processing half the traffic sent their way, and their measured reliability may be as high as $\hat{\rho}_j \approx \frac{1}{2}$ while in truth ${\rho}_j \approx 0$.

\subsection{Threshold approach: $\hat{\beta}_e \in \{0, \frac{1}{2}, 1\}$}
\label{appendix-beta-threshold}

The threshold approach aims to: \textit{(i)} eliminate the unfair degradation of measured reliability for reliable nodes due to successors being offline or congested, or predecessors being consistently faulty;  
while \textit{(ii)} still symmetrically penalizing malicious nodes that conduct selective dropping attacks to degrade the estimated reliability of honest, reliable predecessors or successors. This is achieved by assigning values to $\hat{\beta}_e$ that take into account node reliability across its whole set of incoming or outgoing edges. The reliability $\hat{\rho}_e$ of each edge is computed with Eq.~\ref{eq:rho_e_hat}. 

\noindent
\textbf{Median incoming and outgoing reliability.}
Given $\hat{\rho}_{e} = \frac{{s}^*_e}{{s}^*_e + {d}^*_e}$ for each edge $e$, 
we denote by $\bar{\rho}^{\mathsf{in}}_j =$ med$\{\hat{\rho}_{(i,j)}$ for $i \in P(j)\}$ the weighted median of \textit{incoming} link reliability for node $j$, and by $\bar{\rho}^{\mathsf{out}}_j =$ med$\{\hat{\rho}_{(j,k)}$ for $k \in S(j)\}$ node $j$'s median \textit{outgoing} link reliability. 
The node \textit{weight} in the median computation is given by the share (relative to the full layer) of incoming or outgoing traffic routed by that node. In the case of links between mixnet layers, the node weight is determined by the routing policy. 
All the nodes in a layer count the same weight $\omega^{\mathsf{out}}_i =\omega^{\mathsf{in}}_j =  \frac{1}{W}$ when the routing policy is uniform. 
This is not necessarily the case for gateways, whose weight is instead proportional to their share of sent measurement traffic received by mix nodes in the first layer (i.e., $\omega^{\mathsf{out}}_g = \frac{\sum_i {s}^*_{(g,i)}}{\sum_x \sum_i {s}^*_{(x,i)}}$ when computing $\bar{\rho}^{\mathsf{in}}_i$ for $i \in S(g)$), or received measurements from mix nodes in the last layer (i.e., $\omega^{\mathsf{in}}_g = \frac{\sum_k {s}^*_{(k,g)}}{\sum_x \sum_k {s}^*_{(k,x)}}$ when computing $\bar{\rho}^{\mathsf{out}}_k$ for $k \in P(g)$). %\claudia{@aggelos: please check if this (gateway as successor of last layer mix node) is in/consistent with definitions in sect 2} \aggelos{it is ok based on the latest edits}

\noindent
\textbf{Threshold criteria.}
Let $\bar{\tau}$ be a threshold for the median reliability that nodes are expected to achieve at a minimum when participating in the mixnet (we use $\bar{\tau}=0.99$ in our experimental evaluation).
The $\hat{\beta}_e$ is set per edge $e=(i,j)$ and it takes values in $\{0,\frac{1}{2}, 1\}$, according to the following criteria: 

%\begin{itemize}

 %   \item
 \noindent 
 $\bullet \,\, $
 $\hat{\beta}_{(i,j)} = 1$: 
    if $\bar{\rho}^{\mathsf{in}}_j < \bar{\tau}$ and $\bar{\rho}^{\mathsf{out}}_i \geq \bar{\tau}$ ($j$ has sub-par median incoming reliability $\bar{\rho}^{\mathsf{in}}_j$ while $i$ has above-threshold outgoing median reliability $\bar{\rho}^{\mathsf{out}}_i$). If $\sum_i {s}_{(i,j)}^* = 0$ then $\hat{\beta}_{(i,j)} = 1$ for any value of $\bar{\rho}^{\mathsf{out}}_i$.

 \noindent 
 $\bullet \,\, $
 $\hat{\beta}_{(i,j)} = 0$: if $\bar{\rho}^{\mathsf{in}}_j \geq \bar{\tau}$ and $\bar{\rho}^{\mathsf{out}}_i < \bar{\tau}$ ($j$ has above-threshold median incoming reliability while $i$ is showing low reliability in more than half of its outgoing links).

 \noindent 
 $\bullet \,\, $
 $\hat{\beta}_{(i,j)} = \frac{1}{2}$: if $\bar{\rho}^{\mathsf{out}}_i, \bar{\rho}^{\mathsf{in}}_j \geq \bar{\tau}$ (both $i$ and $j$ have above-threshold median reliability), or    
    $\bar{\rho}^{\mathsf{out}}_i, \bar{\rho}^{\mathsf{in}}_j < \bar{\tau}$ (both $i$ and $j$ have sub-par median reliability).  
    
%\end{itemize}

Using median values $\bar{\rho}^{\mathsf{in}}_j$ and  $\bar{\rho}^{\mathsf{out}}_j$ to infer whether a node is performing adequately across the board makes the approach robust to adversarial settings, up to the point where the adversary controls enough resources to affect median values. In the case of uniform routing and $W$ nodes per mixnet layer this means controlling $\frac{W}{2}$ nodes in a layer; while in the case of gateways it involves controlling gateways that combined send half the total client traffic or that receive half of all the measurement packets.

\section{Mixnet Simulator}
\label{appendix-simulator}

This section provides a description of the simulator we implemented\footnote{Available at: \url{https://github.com/claudia-diaz/artifact-decentralized-measurements-EuroSP2026}.} 
%\footnote{The code and datasets can be made available upon request during submission and will be published with the paper.} 
and used to conduct an empirical analysis of the proposed reliability estimation protocol.
The scripts are in Python and use the SimPy library to handle events such as packet creation, packet sending, receiving and dropping. %We have two versions of the simulator that use the same core functions, one configured for reliability estimation and another for free riding detection experiments. 
This section provides extensive details on the features, parameters and experimental setups used in to obtain the results presented in previous sections, as well as additional experiments excluded from the main body due to page limit. 

\subsection{Clients and Encoded Packets}

The simulator runs clients that encode and send packets throughout a simulated epoch of one hour. Packet generation by clients follows a Poisson process with a configurable rate of $\lambda$ packets per second. Each encoded packet is a measurement packet with (globally configurable) probability $p_\mathsf{lot}$. The rest are data packets addressed to other (randomly chosen) clients or cover traffic that clients send to themselves. 
Packets' routes are selected according to a uniform routing policy, considering a mixnet with $L=3$ layers and $W=80$ mix nodes per layer, as well as $W_G=80$ gateways.%\footnote{To make the simulations as realistic as possible, we select network and mixing delay parameters similar to those in the deployed Nym network.} %Packets may or may not be acknowledged, depending on the configuration. If they are acknowledged  their full route includes ten links, five from the sending client to the packet's destination gateway (via the sending client's gateway and three mix nodes) and another five for the packet's acknowledgment to reach back the client. If packets are not acknowledged, they only have a forward route of five hops. We use acknowledged packets in the reliability estimation experiments and (to speed up simulations) non-acknowledged packets in the free riding detection experiments. 
We consider that legitimate client traffic is uniformly distributed over all gateways and also addressed in equal measure to recipients in all gateways. The encoded delay per packet per mix node is exponentially distributed with average $50$ms, and we additionally consider $40$ms of transmission time in each link and $2$ms of packet processing time at the gateways. 

%The simulator runs a configurable number of client instances. Independently of each other, these clients randomly switch between being online and offline (for exponentially long periods), and while online they can switch (after exponentially long periods) between a baseline encoding rate and a high encoding rate. This allows testing scenarios with stable traffic rates as well as scenarios where traffic rates are highly variable over time. \claudia{this is implemented but not really used in the final experiments}
%The current version of the scripts focuses on evaluating the reliability estimation approach and does not consider injected traffic. 

\subsection{Reliable, Unreliable and Malicious Nodes}

In the simulations we treat gateways and mix nodes equally in terms of reliability, failure modes and malicious behaviour, and thus refer to all $80$ gateways and $240$ mix nodes indistinctly as `nodes' that receive packets from their predecessors and forward them to their successors, sometimes dropping packets if they are adversarial or not fully reliable.
%and sometimes adding packets if they are free riders. 
Packets may be dropped by a node before or after the packet tag is stored in the node's tag-commitment. If a packet is dropped \textit{before} storing, it is considered that the drop happened in the node's incoming link. If the drop happens \textit{after} storing then it is accounted for in the node's outgoing link.
Nodes can have various types of behaviours: 
%\begin{itemize}

 %   \item
 \noindent 
 $\bullet \,\, $
\textit{Reliable} nodes are honest and function perfectly: they are always online and correctly process all the received packets without dropping any of them.

 %   \item
 \noindent 
 $\bullet \,\, $
 \textit{Unreliable} or \textit{faulty} nodes are non-adversarial but subject to failures that cause packet drops. The possible failures include: (a) going offline for some (exponentially distributed) period of time, dropping any packets that were inside the node at the moment of going offline, as well as failing to receive new packets while offline; (b) becoming congested due to having a limited throughput and dropping packets on incoming links when the incoming packet rate exceeds the node's throughput; (c) suffer from random glitches that cause packet drops in incoming or outgoing links with a certain probability.

 %   \item
 \noindent 
 $\bullet \,\, $
 \textit{Malicious} nodes are always online and have high throughput but they \textit{purposefully} drop all or a fraction of packets in links with \textit{selected} predecessor or successor targets, in order to degrade the reliability estimates $\hat{\rho}_j$ for those targets.  
    
    %\item \textit{Free Rider} nodes add their own traffic with a (configurable) combination of probabilistic packet \textit{substitution} and \textit{injection}. When forwarding a packet received from a predecessor, the free rider replaces the received packet by its own packet with a certain probability of substitution, and with a probability of injection the free rider generates and sends an additional packet. We consider that the routes of free riding packets follow the same (uniform) routing policy as legitimate traffic, in order for free riders to maximize the amount of free riding traffic they inject (staying under the radar in as many links as possible).

%\end{itemize}

%\claudia{Parameters: capacity per node, prob glitch, online-offline mean periods, nr of adversaries/targets + prob of attack}

\subsection{Simulation Outputs} 

Once the epoch finalizes, the simulator logs in .csv files the outputs of the sampling protocol, including the ${s}^{*}_e$, ${d}^{*}_e$ counts per link $e$ as well as $\hat{\rho}_j$, $\bar{\rho}^{\mathsf{in}}_j$,  $\bar{\rho}^{\mathsf{out}}_j$  
estimates for every node $j$. 
In addition, the simulator logs packet counts for all edges and nodes that would not be available in an actual deployment -- but that are available in a simulated environment and useful for evaluation purposes. 
This includes ${s}_e$, ${d}_e$, 
${\rho}_e$, and ${\beta}_{e}$ for all edges as well as ${\rho}_j$ for all nodes and gateways. 
We evaluate the proposed reliability estimation protocols by comparing the $\hat{\rho}_j$ output by the protocol to the underlying ground truth reliability ${\rho}_j$, both in settings where packet drops are due to random failures and %adversarial
settings where packet drops are malicious.

\subsection{Simulation Runtime}

The runtime of a simulation is proportional to the number of packets encoded and routed in the full simulation. Each packet triggers a number of events, first as it is created by a client and then every time it is sent or received by an intermediary routing (or dropping) the packet. 
We perform simulations with a total of encoded packets between $2$ million and $200$ million, which require, respectively, between $15$-$20$ minutes and $30$-$36$ hours to complete. Simulations are CPU intensive (each simulation fully utilizes a CPU) but require little memory. 
%We use Jupyter Notebook scripts to read the output .csv files of multiple simulations and display the aggregate results as graphs shown in the results.

\subsection{Experimental Setup: Honest but Unreliable Setting} 
\label{exp-setup-reliability}

In this setup we simulate scenarios where in each layer $40$ nodes (half the layer) are reliable and the other $40$ are (potentially) unreliable in various ways. To test the effect of different modes and degrees of failure we consider:

 %   \item
 \noindent 
 $\bullet \,\, $
 $32$ unreliable nodes per layer ($40\%$ of the layer) \textit{may} go \textit{offline} during the epoch. Each of these nodes toggles between being online and offline for exponentially distributed lengths of time, with average $90$ minutes online and $10$ minutes offline; i.e., on average, these nodes are online $90\%$ of the time. %Within a one-hour epoch however, about half of these nodes stay online the entire epoch and in practice behave as reliable, with the other half going offline for at least part of the epoch. 

 %   \item
 \noindent 
 $\bullet \,\, $
 $4$ unreliable nodes per layer ($5\%$ of the layer) have \textit{limited throughput}, with one of them having a throughput equal to the average per-node incoming packet rate (and thus only dropping some packets when traffic is higher than average), and the other three having throughput that is $\frac{1}{2}$, $\frac{1}{4}$, and $\frac{1}{8}$ of the average incoming packet rate (and thus dropping, respectively, about $\frac{1}{2}$, $\frac{3}{4}$, and $\frac{7}{8}$ of incoming packets).

 %   \item
 \noindent 
 $\bullet \,\, $
 The final $4$ unreliable nodes per layer ($5\%$ of the layer) \textit{randomly drop} some packets. One of these nodes drops $1\%$ of the incoming packets, another drops $1\%$ of outgoing packets, a third drops $20\%$ of incoming packets and the last drops $20\%$ of outgoing packets.

%Note that of the $32$ nodes that \textit{may} go offline during the epoch, not all do, and thus some of these nodes act reliably during the epoch. In practice about two thirds of all gateways and nodes are reliable in each simulation, while the remaining third drops at least one packet.  

We set the probability of a packet being a measurement to $1\%$, i.e., $p_\mathsf{lot}=0.01$. We vary the total amount of packets encoded by clients to study the node reliability estimation accuracy relative to the number of measurement samples %taken 
throughout the epoch. The results are shown in Figure~\ref{fig:perf-reliability}.

\subsection{Experimental Setup: Adversarial Settings} 
\label{appendix-setup-adversarial}

To simulate adversarial settings we consider scenarios with $A=\{1, 2, 4, 8, 16, 32, 64\}$ adversaries and $T=\{1, 2, 4, 8, 16, 32, 64\}$ targets. We only consider node assignments to layers that are most favorable to the adversary, i.e., where \textit{all} $A$ adversarial nodes are in layers \textit{adjacent} to \textit{all} the $T$ targets, either as predecessors, as successors, or both. In cases where all $A$ nodes are in a single layer, the $T$ nodes may be all in the preceding layer, all in the succeeding layer or, for $T>1$, distributed as $\frac{T}{2}$ in the preceding and $\frac{T}{2}$ in the succeeding layers. In cases where the $A$ nodes are distributed among two layers, we consider that all $T$ targets are in the layer between the adversaries, with $\frac{A}{2}$ as predecessors and the other $\frac{A}{2}$ as successors of the $T$ targets. For $64$ adversaries or targets we only consider scenarios where they are split over two layers (rather than all in one layer). 
We consider that the $A$ adversaries maximize the attack impact by dropping \textit{all} (rather than just some) of the packets in the edges shared with the $T$ targets -- while not dropping \textit{any} packets on edges shared with non-targets. 
Targets are reliable nodes that do not drop any packets and thus have a true reliability of $100\%$. For the remaining `vanilla' nodes, which are non-adversarial and non-targets, we consider two cases: one where they are all reliable, and another where $30\%$ of the nodes in each layer are potentially unreliable (results reported in Section~\ref{sec:perf-adversarial}). The second case uses a  setup almost identical to the one described in Section~\ref{exp-setup-reliability}, with the only difference being that only $16$ (rather than $32$) unreliable nodes per layer may go offline during the epoch. 
% \claudia{NOTE: we could (should?) address this and have a second graph in the results section showing results for the attack when some of the vanilla (non-target, non-adversarial) nodes are \textbf{unreliable}.}
%In order to evaluate a large number of simulated scenarios within a reasonable time frame, we set $p_\mathsf{lot}=1$ in these simulations, meaning that in practice we exclude non-measurement packets from the simulation. This speeds up by $100$x the time needed to obtain experimental results without sacrificing accuracy of evaluation, since both targets and adversaries are considered to be reliable in the absence of attack and their measured reliability loss is thus compared to having a reliability of $100\%$. Thus, simulating non-measurement packets does not add any useful information to the output while multiplying simulation time by 100.
We consider scenarios with a total of $2$ million measurement packets per epoch of one hour.

\section{Protocol overhead evaluation}
\label{sec:eval-overhead}
%\claudia{==== begin notes ====}
%Blockchain storage (long-term): 
%\begin{itemize}
    %\item beyond the scope of our protocol: global parameters, client subscriptions, spent credentials, public keys of gateways / nodes, assignment of nodes to mixnet layers, node staking and rewards, blocklisted (found to be malicious) nodes, ...
%    \item gateways: per-credential spent counters (keep track of credentials' unused budget across epochs), 
  %  \item results: per-node reliability score $\hat{\rho}_j$ (consensus among validators who locally compute the value from the broadcast info). 
 %   \item \claudia{@@aggelos: any other proofs or info that need to be permanently stored in the blockchain?}
%\end{itemize}
%Ephemeral storage / broadcasting (info can be forgotten after a few hours): 
%\begin{itemize}
    %\item Per-packet openings: size opening x nr of openings (2 million?)
%    \item Bloom filter: size filter x nr of nodes (+ gateways with smaller filter)
    %\item Proofs of no-skipping: size proofs x nr of proofs 
    %\item 
%\end{itemize}
%\claudia{==== end notes ====}

We evaluate the overhead introduced by the proposed protocols in terms of the data that must be broadcast per epoch, the data that must be persisted long-term on a blockchain, and the computational overhead of VRF-based routing. 
%In transit through the mixnet, processing packets only requires a single exponentiation per hop for determining the next hop in the route, which is already done by Sphinx as described in Appendix~\ref{sphinx}. 
%Thus, no additional per-packet computational overhead is needed in concrete implementations of our reliability estimation scheme. 
% 
We assume that several variables required by our protocols are already publicly available for the baseline operation of the mixnet, even before introducing any reliability measurement mechanisms. This includes: the mixnet topology definition, per-epoch node-to-layer assignments, node public keys (used for signing broadcast data and decrypting packets), node reward information, client subscriptions, and spent credentials (corresponding to step \textbf{B.1} in Fig.~\ref{fig:overall-system}). We therefore focus solely on the additional overhead incurred by the reliability estimation protocol, as summarized in Table~\ref{tab:overhead}. 

We consider a setup in which $100$ million packets are routed per epoch, of which $1$ million ($1\%$) are measurement packets. The mixnet comprises $80$ gateways and $240$ mix nodes, divided across three layers of $80$ nodes each. 
The communication complexity when using Merkle trees as tag-commitments is discussed in Sect.~\ref{link-reliability-honest}. 
Here we assume Bloom filters are used for tag-commitments, thereby making steps B.4 and B.6 in Fig.~\ref{fig:overall-system} unnecessary.
The data that must be broadcast in each epoch to compute reliability scores includes:

\noindent
\emph{Bloom filters}: Each node must broadcast its tag-commitment at the end of the epoch (step \textbf{B.2} in Fig.~\ref{fig:overall-system}). In the considered setup, a mix node is expected to route $1/80$ of the total traffic, amounting to $1.25$ million packets. Assuming a false positive rate of $10^{-5}$, this results in a Bloom filter of approximately $3.5$ MB in size.\footnote{We use Bloom filter calculators available at \url{https://hur.st/bloomfilter/}} With $240$ mix nodes, the total Bloom filter data adds up to $840$ MB. The $80$ gateways can use smaller Bloom filters, as they only need to record received measurement packets. Assuming a %false positive rate of $10^{-5}$ and a 
maximum of $100,000$ received measurements per gateway, each filter would be about $300$ KB, totaling $24$ MB across all gateways.  

\noindent 
\emph{ Merkle Trees}: Each node must broadcast a tag-commitment at the end of the epoch (step \textbf{B.2} in  Fig.~\ref{fig:overall-system}) which is merely 32 bytes (the Merkle tree root) assuming SHA256 as the underlying hash function. Subsequently, each node should open the tags that correspond to the measurement packets 
(step \textbf{B.4} in  Fig.~\ref{fig:overall-system}). Assuming a maximum of 100,000 measurements, for mix nodes, this would require  17 hashes per membership proof resulting in a total of  $55$ MB of data per node.

\noindent
\emph{Per-packet openings}: Gateways broadcast an opening per measurement packet (step \textbf{B.3} in Fig.~\ref{fig:overall-system}), each of size 388 bytes. For $1$ million measurement packets, this adds up to 388 MB of broadcast data.

\noindent
\emph{Proofs of no-skipping}: Each opening for a non-measurement position is 132 bytes. Given a total of 99 million non-measurement packets per epoch, we consider an adversarial gateway that hides one measurement packet. By requiring proofs for  $99\cdot 10^6 \ln (1-.01)^{-1} \approx 10^6 $ non measurement positions,  the probability of detecting the omission is at least $1\%$. The resulting data overhead is $132$ MB, and such a gateway would be caught on average within $100$ epochs.

We consider that the computed reliability scores for all nodes and gateways are stored persistently in the blockchain to act as a historical record of their measured reliability. 
Assuming a precision of four decimal places, this requires $2$ bytes per node and gateway, resulting in a total of $640$ bytes per epoch for $240$ nodes and $80$ gateways. 
Additional protocol parameters that must be stored on-chain, such as $p_\mathbf{lot}$ and $\bar{\tau}$, occupy only a few extra bytes and do not require updates every epoch.

%\claudia{@@aggelos: any other proofs or info that need to be broadcast or stored permanently in the blockchain?}

\begin{table}
    \centering
    \begin{tabular}{lccc}
        \hline
         \textbf{Broadcast} & \textbf{Per item} & \textbf{Amount} & \textbf{Total data} \\
         Packet openings & 388 B & 1 million & 388 MB \\
         Bloom filters mix nodes & 3.5 MB & 240 & 840 MB \\
         Bloom filters gateways & 300 KB & 80 & 24 MB \\
         No-skipping proofs & 132 B & 1 million & 132 MB \\
        \hline
         %\textbf{Long-term Storage} &  & \\
         %Node reliability scores & 2 B & 320 & 640 B \\
         %\claudia{anything else?} & & & \\
         %& & & \\
         %\hline
    \end{tabular}
    \caption{Overhead broadcast per epoch}
    \label{tab:overhead}
\end{table}

In terms of computational overhead, based on public benchmarks from Nym for Sphinx packet processing,\footnote{See \url{https://github.com/nymtech/sphinx-bench}.} the per-hop latency on a standard Linux server is approximately $0.32$ms. We observe that VRF-based routing adds at most the equivalent of a single Sphinx processing step, keeping the total per-hop latency below $1$ms. 
Moreover, combining our scheme with Sphinx allows for further optimizations that can further reduce this overhead.
Each packet requires three VRF evaluations, but these can be performed in advance during a preprocessing phase, so they do not affect transmission latency. 
If packet creation occurs in real time rather than via preprocessing, the three VRF operations would introduce approximately $1$ms of additional computation. With simple optimizations, this cost can be reduced to a single VRF evaluation per packet. 
%extrapolating from Sphinx processing  this will require at most $1$ ms pre-processing (given that the dominant computational overhead in Sphinx comes from scalar multiplication in the curve --- the same operation we need for the VRF). 
Aside from these negligible increases in latency introduced by VRF-based routing, our reliability estimation protocols do not affect real-time packet forwarding, as all computationally intensive steps are performed \textit{a posteriori}---after the packets for an epoch have been delivered---rather than during packet transmission. 
The protocols are designed to complete within a single epoch, which is easily achievable for epoch durations of one hour (as in the current Nym mixnet) or longer.

\end{document}